\documentclass[onecolumn, draftclsnofoot, 12pt]{IEEEtran}
\usepackage{amsmath}
\usepackage{cite}
\usepackage{amssymb}
\usepackage{amsfonts}
\usepackage{algorithm}
\usepackage{algpseudocode}
\usepackage{dsfont}
\usepackage{graphicx}
\usepackage{epsfig}
\usepackage{subfigure}
\usepackage{psfrag}
\usepackage{xcolor}
\usepackage{url}
\usepackage[colorlinks,linkcolor=black,urlcolor=black,anchorcolor=black,citecolor=black,hyperfootnotes=true]{hyperref}

\newtheorem{theorem}{Theorem}

\title{{A Heterogeneous 6G Networked Sensing Architecture with Active and Passive Anchors}
\thanks{Q. Wang, L. Liu, S. Zhang, and C. M. Lau are with the Department of Electronic and Information Engineering, The Hong Kong Polytechnic University, Hong Kong SAR, China (e-mails: qipeng.wang@connect.polyu.hk, \{liang-eie.liu, shuowen.zhang, francis-cm.lau\}@polyu.edu.hk).}
\thanks{B. Di is with the Department of Electronics, Peking University, Beijing 100871, China (email: diboya@pku.edu.cn).}
\thanks{The materials in this paper have been presented in part at the IEEE Global Communications Conference, December 2022 \cite{globecom22}.}}
\author{\IEEEauthorblockN{Qipeng Wang, Liang Liu, Shuowen Zhang, Boya Di, and Francis C. M. Lau}}

\begin{document}
		\maketitle \thispagestyle{empty} \vspace{-30pt}
\begin{abstract}
In the future 6G integrated sensing and communication (ISAC) cellular systems, networked sensing is a promising technique that can leverage the cooperation among the base stations (BSs) to perform high-resolution localization. However, a dense deployment of BSs to fully reap the networked sensing gain is not a cost-efficient solution in practice. Motivated by the
advance in the intelligent reflecting surface (IRS) technology for 6G communication, this paper examines the feasibility of deploying the low-cost IRSs to enhance the anchor density for networked sensing. Specifically, we propose a novel heterogeneous networked sensing architecture, which consists of both the active anchors, i.e., the BSs, and the passive anchors, i.e., the IRSs. Under this framework, the BSs emit the orthogonal frequency division multiplexing (OFDM) communication signals in the downlink for localizing the targets based on their echoes reflected via/not via the IRSs. However, there are two challenges for using passive anchors in localization. First, it is impossible to utilize the round-trip signal between a passive IRS and a passive target for estimating their distance. Second, before localizing a target, we do not know which IRS is closest to it and serves as its anchor. In this paper, we show that the distance between a target and its associated IRS can be indirectly estimated based on the length of the BS-target-BS path and the BS-target-IRS-BS path. Moreover, we propose an efficient data association method to match each target to its associated IRS. Numerical results are given to validate the feasibility and effectiveness of our proposed heterogeneous networked sensing architecture with both active and passive anchors.
\end{abstract}
\section{Introduction}
\subsection{Motivation}
As the two most important applications of wireless technologies, communication and sensing have traditionally been designed and implemented separately using their own hardware and spectrum. Recently, there is a trend in both academia and industry towards integrated sensing and communication (ISAC) in the future 6G cellular networks \cite{isac_survey1, isac_survey2, isac_survey3, Tan21}, where the base stations (BSs) can emit wireless signals not only to convey information to communication users, but also to sense the environment with high resolution, thanks to the wide bandwidth at the millimeter wave (mmWave) band. In the 6G-based ISAC realm, the cellular communication technique is quite mature. However, it remains an open problem in how to leverage the 6G network for achieving the best sensing performance.

In general, sensing can be classified into device-based sensing, which aims to localize active targets with communication capabilities, e.g., user equipments, based on the one-way signal propagated between the anchor and each target, and device-free sensing, which is also able to localize passive targets without communication capabilities, based on the echo signals reflected by these targets. This paper considers the device-free sensing technique, due to its generality to localize different kinds of targets. In particular, motivated by the cooperative communication technique such as cloud radio access network (C-RAN) and coordinated beamforming (CoMP) where multiple BSs jointly encode/decode the user messages, this paper focuses on the networked device-free sensing technique \cite{dvc}, where multiple BSs can share their sensing information obtained from the echo signals to jointly localize the targets. When multiple BSs cooperatively localize multiple passive targets, the \emph{data association issue} \cite{dvc,Mahler07} arises, because it is hard for all the BSs to pick up the echoes belonging to the same target. The recent work \cite{dvc} showed that under the range-based trilateration method, data association will not fundamentally affect the performance of networked sensing. Moreover, an efficient data association algorithm was proposed such that all the BSs can identify each target's echo signals for localizing it.

\begin{figure}[t]
	\centering
	\includegraphics[width=10cm]{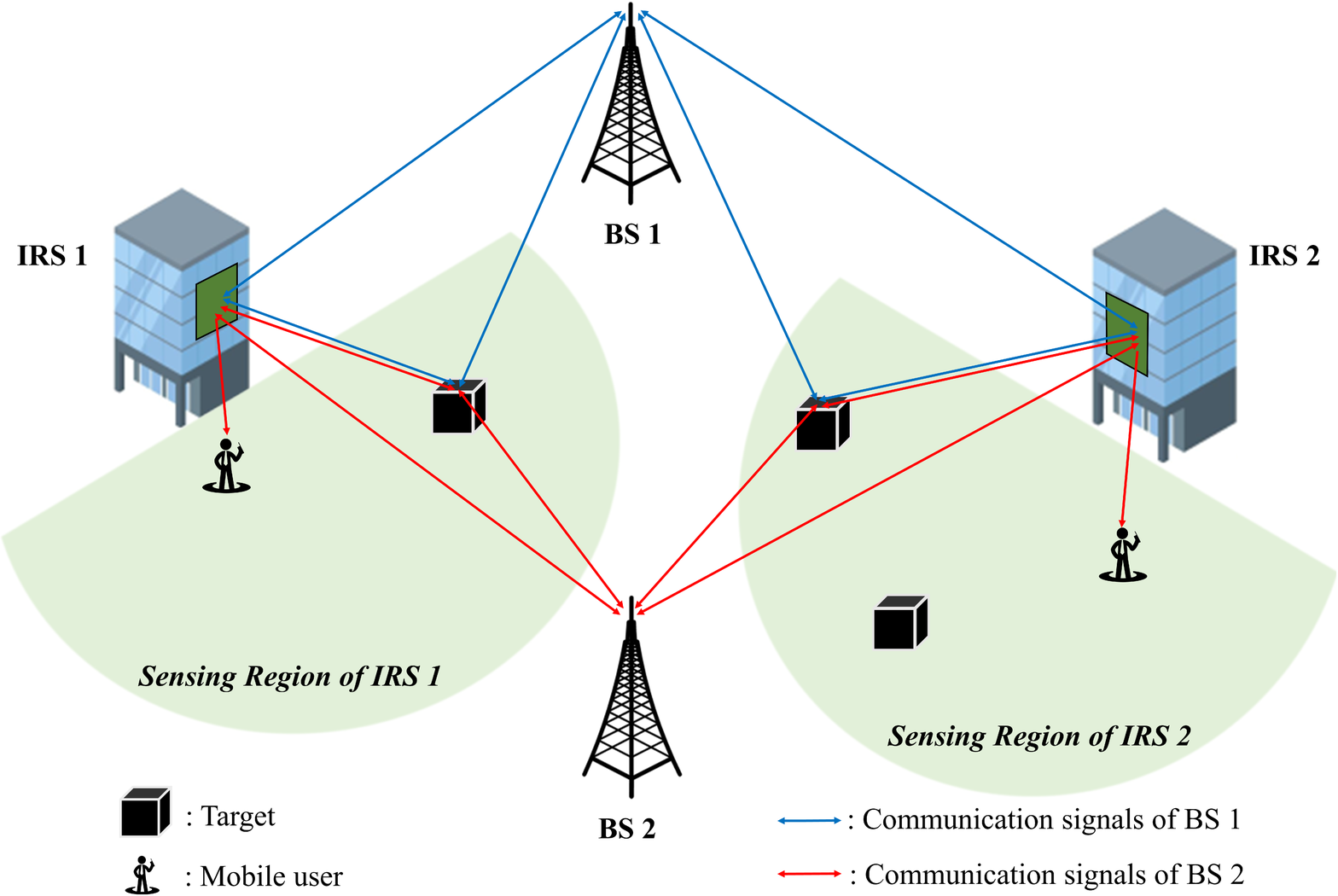}
        \caption{An example of the heterogeneous architecture for networked sensing with two active BSs and two passive IRSs.} \label{fig:system}
\vspace{-5mm}
\end{figure}

In general, the networked sensing performance improves with the number of anchors. However, it is not cost-efficient to densely deploy the BSs to reap the joint localization gain. Recently, there has been a flurry of research activities in using intelligent reflecting surfaces (IRSs) to enhance the network coverage and capacity in the 6G era \cite{irs_survey1,irs_survey2}. Because IRSs are made by passive elements, they can be densely deployed with low cost and low energy consumption \cite{irs_survey1,irs_survey2}. Therefore, in this paper, we propose a novel heterogeneous networked sensing architecture that consists of both the active anchors, i.e., BSs, and the passive anchors, i.e., IRSs, to enhance the anchor density and improve the networked sensing gain, as shown in Fig. \ref{fig:system}. Our goal is to illustrate how to leverage hybrid anchors under the above architecture to localize multiple targets, even if some anchors cannot actively transmit/receive the radio signals.

\subsection{Prior Work}
The feasibility and effectiveness for using IRSs in 6G-based sensing have been recently studied in the literature \cite{irs_loc_mag,irs_loc_mag2,emil22}. Under the device-based sensing setup, various works have investigated how to utilize the IRSs as anchors for localization. Specifically, \cite{Dardari22} considered a range-based localization system consisting of a BS, a large IRS with multiple tiles, and a user equipment to be localized. By viewing the BS and IRS tiles as anchors with known locations, an efficient two-step positioning scheme was proposed, where the distances from the user to the BS and to the IRS tiles are first estimated based on the one-way signals between the BS and the user via/not via the IRS, and the location of the user is then estimated based on the above range information. Moreover, because an IRS can be viewed as a linear array, the angle information between the user equipment and the IRS can also be estimated and utilized for localization. Along this line, \cite{aoa} considered an angle-based localization system consisting of a BS, multiple IRSs, and a user equipment. A two-step algorithm was proposed to solve the maximum likelihood (ML) problem for estimating the angle information, where an exhaustive search method is first implemented over a discrete grid to find some good initial points, and a gradient decent search method is then applied to find a better angle solution. Then, the user can be localized based on its angles to different anchors, i.e., the BS and the IRSs. Last, \cite{irs_aoa_toa1} and \cite{irs_aoa_toa2} proposed to utilize both the range and angle information to localize the user equipment via the IRS. Similar to \cite{aoa}, both the range and the angle between the user equipment and the IRS are estimated based on a two-step approach, where a coarse estimation is first made via the exhaustive search method over a discrete grid, and a better solution is then made by refining the coarse estimation. On the other hand, under the device-free sensing setup, several works consider the usage of IRSs to enhance the strength of the echo signals from the target for better detection performance \cite{irs_dvc1,irs_dvc2}. In these works, various algorithms were proposed to optimize the IRS reflecting coefficients so as to maximize the target detection probability.

To summarize, under the device-free sensing setup, how to utilize the IRSs as anchors to localize multiple targets is still an open problem. Note that although the IRSs have been used as anchors in device-based sensing \cite{Dardari22,aoa,irs_aoa_toa1,irs_aoa_toa2}, these algorithms cannot be applied for our considered networked device-free sensing architecture, due to the following reasons. First, under device-based sensing, different user equipments can independently estimate their angles to the anchors based on their received signals. The complexity for each user to perform one-dimension search as in \cite{Dardari22,aoa,irs_aoa_toa1,irs_aoa_toa2} is reasonable. However, under device-based sensing, the BS's received signal is a superposition of the echo signals from multiple targets. In this case, the angle information of all the targets have to be jointly estimated. If we follow the approach in \cite{Dardari22,aoa,irs_aoa_toa1,irs_aoa_toa2}, multi-dimension search is needed to roughly estimate different targets' angle information, which is of high complexity. Second, different from device-based sensing where the signals to/from different user equipments have different signatures, under device-free sensing, each BS's signal is reflected back by multiple targets with the same signature. This leads to the data association issue \cite{dvc,Mahler07} for networked device-free sensing, which is not solved in \cite{Dardari22,aoa,irs_aoa_toa1,irs_aoa_toa2}.

\subsection{Main Contributions}
Under our proposed heterogeneous networked sensing architecture, this paper considers a 6G-based sensing network that consists of two active BSs, multiple passive IRSs, and multiple passive targets, as shown in Fig. \ref{fig:system}. In the downlink, the BSs transmit the orthogonal frequency division multiplexing (OFDM) signals, while these signals can be reflected back to the BSs via the BS-target-BS path and the BS-target-IRS-BS path. In this paper, we assume that the distance between any two IRSs is sufficiently long such that each IRS covers a separate sensing region. In other words, each target is merely in the coverage region of one IRS that is closest to it, while the other IRSs are too far away to reflect signals from/to this target. Under such a setup, the two BSs and the associated IRS can serve as three anchors to localize any target via the trilateration method, if their distances to this target can be measured. Based on this observation, we adopt a two-phase localization protocol to localize the targets based on their echoes \cite{dvc}. In the first phase, we apply the OFDM channel estimation technique to obtain the delay (thus the range) of each BS-target-BS path and  BS-target-IRS-BS path, while in the second phase, the targets' locations are estimated based on the above range information. However, two challenges arise in the above protocol, due to the passive anchors in the system. Challenge I: Since both IRSs and targets are passive, it is impossible to utilize the round-trip signals between a target and its associated IRS for estimating their distance. Challenge II: Before localizing a target, we do not know which IRS is closest to it and serves as its anchor. This paper aims to tackle the above challenges via the advanced signal processing technique so as to enable our proposed heterogeneous networked sensing architecture. The main contributions of this paper are summarized as follows.

First, we manage to tackle Challenge I under our proposed heterogeneous networked sensing architecture. The key idea is that although the distance from a passive target to its associated passive IRS cannot be directly obtained based on the round-trip signals between them, it can be measured indirectly in our considered system. Specifically, as shown in Fig. \ref{fig:system}, via utilizing our proposed OFDM channel estimation method, the active BSs can estimate the length of the BS-target-BS path and that of the BS-target-IRS-BS path based on the echoes passing through these links. Moreover, the distances between the BSs and the IRSs are known. Therefore, for any target and its associated IRS (assuming that such an association is known), their distance can be calculated by subtracting the BS-target distance and the BS-IRS distance from the length of the BS-target-IRS-BS path.

Second, we manage to tackle Challenge II under our proposed heterogeneous networked sensing architecture. The key idea is that after we associate one IRS to one target, BS $1$ can estimate the distance between this target and this IRS based on its measured length of the BS $1$ to target to BS $1$ path and that of the BS $1$ to target to its associated IRS to BS $1$ path, while BS $2$ can estimate this distance similarly. If the matching between the target and the IRS is correct, the estimations of the distance between this target and this IRS made by the two BSs should be very close to each other. We reveal that this property can be leveraged to tackle the target-IRS association issue. Specifically, in the ideal case that the ranges of all the BS-target-BS links and the BS-target-IRS-BS can be perfectly estimated, we rigorously show that we are able to take advantage of the above property for finding the associated IRSs for all the targets correctly with probability one. Moreover, in the practical case with imperfect range estimations, we propose an efficient method based on the above property to determine the target-IRS association solution.

Third, via sufficient numerical examples, we show that given the same number of anchors, our proposed heterogenous networked sensing architecture with both active and passive anchors achieves similar or sometimes better performance compared to the traditional networked sensing architecture merely with active anchors. This is because by leveraging the estimations of the target-IRS distance made by two BSs, we can significantly reduce the size of the feasible data association set such that the accurate localization solution can be found much more easily. This verifies the feasibility and the effectiveness to employ low-cost IRSs as anchors to improve the networked sensing gain. In other words, in the future 6G network, passive IRSs will not only enhance the channel quality for high-speed communication, but also serve as anchors for ultra-accurate localization.

\subsection{Organization}
The rest of the paper is organized as follows. Section \ref{sec:system_model} introduces the system model. Section \ref{sec:protocol} presents the two-phase localization protocol. Sections \ref{sec:range_est} and \ref{sec:da_loc} introduce the range estimation phase and the localization phase of this protocol, respectively. Numerical results are provided in Section \ref{sec:numerical}. Finally, Section \ref{sec:conclusion} concludes this paper.

\section{System Model}\label{sec:system_model}
We consider an IRS-assisted OFDM-based ISAC system as illustrated in Fig. \ref{fig:system}, which consists of two BSs, $R\geq 1$ IRSs, $S$ mobile users to be served, and $K$ targets to be localized. Each BS $m$ is equipped with one transmit antenna, one communication receive antenna, and one sensing receive antenna to perform ISAC. Moreover, each IRS $r$ is equipped with $I_r$ reflecting elements, $\forall r$. In the considered system, the BSs emit the OFDM signals in the downlink for conveying information to the mobile users and localizing the targets based on their echoes, where the IRSs can simultaneously improve the channel quality of the mobile users and serve as anchors with known locations to help localize the targets. Since IRS-assisted communication has been widely studied in the literature \cite{irs_survey1,irs_survey2}, this paper focuses on IRS-assisted sensing to realize ISAC in the considered OFDM-based system.

In this paper, we assume that the IRSs are deployed sufficiently far away from each other such that their sensing regions for the targets are non-overlapping due to the severe path loss, as shown in Fig. \ref{fig:system}. Such assumptions are widely made in the literature of multi-IRS-assisted communication \cite{multiirs1,multiirs3}, where different IRSs serve different sets of mobile users. Let $\gamma_k$ denote the index of the IRS that target $k$ is associated with, $\forall k$. Then, merely the echoes reflected from/to IRS $\gamma_k$ to/from target $k$ are strong enough to be detected by the BSs, while those reflected from/to the other IRSs to/from target $k$ are too weak to be detected. In the two-dimensional (2D) Cartesian coordinate system, let $(x_m^{\text{B}},y_m^{\text{B}})$ and $(x_r^{\text{I}},y_r^{\text{I}})$ in meter denote the known locations of the $m$-th BS, $m=1,2$, and the $r$-th IRS, $r=1,\cdots,R$, and $(x_k^{\text{T}},y_k^{\text{T}})$ in meter denote the unknown location of the $k$-th target, $k=1,\cdots,K$, respectively. Then, define
\begin{align}
    d_{m,{k}}^{\text{BT}}&=\sqrt{(x_m^{\text{B}}-x_{k}^{\text{T}})^2+(y_m^{\text{B}}-y_{k}^{\text{T}})^2},\quad\forall m,k, \label{eqn:dis1}\\
    d_{\gamma_k,k}^{\text{IT}}&=\sqrt{(x_{\gamma_k}^{\text{I}}-x_{k}^{\text{T}})^2+(y_{\gamma_k}^{\text{I}}-y_{k}^{\text{T}})^2}, \quad \forall k, \label{eqn:dis2} \\
    d_{m,r}^{\text{BI}}&=\sqrt{(x_m^{\text{B}}-x_r^{\text{I}})^2+(y_m^{\text{B}}-y_r^{\text{I}})^2},\quad\forall m,r, \label{eqn:dis3}
\end{align}in meter as the distance between target $k$ and BS $m$, the distance between target $k$ and its associated IRS $\gamma_k$, and the distance between BS $m$ and IRS $r$, respectively. If $d_{m,{k}}^{\text{BT}}$'s, $m=1,2$, and $d_{\gamma_k,k}^{\text{IT}}$ can be estimated from the echoes reflected by target $k$ not via/via its associated IRS $\gamma_k$, then BS $1$, BS $2$, and IRS $\gamma_k$ can serve as three anchors with known locations to localize target $k$ via the trilateration method based on (\ref{eqn:dis1}) and (\ref{eqn:dis2}), $\forall k$.

To achieve the above goal, the BSs will emit the OFDM signals in the downlink such that the range information can be extracted from the targets' echoes. Let $N$ and $\Delta f$ denote the number of OFDM sub-carriers and the sub-carrier spacing, respectively. Then, the overall channel bandwidth is $B=N\Delta f$ Hz. Moreover, let $\mathcal{N}_1$ and $\mathcal{N}_2$ denote the set of sub-carriers allocated to BS $1$ and BS $2$, respectively. We assume that the two BSs are assigned with orthogonal sub-carriers, i.e., $\mathcal{N}_1 \cap \mathcal{N}_2=\emptyset$ and $\mathcal{N}_1 \cup  \mathcal{N}_2=\{1,\cdots,N\}$, to avoid inter-cell interference \cite{fdd}. Moreover, we use $Q$ to denote the total number of OFDM symbols. In each $q$-th OFDM symbol duration, let $s_{m,n}^{(q)}$ denote the signal of the $m$-th BS at the $n$-th sub-carrier, with $\mathbb{E}[|s_{m,n}^{(q)}|^2]=1$ if $n\in \mathcal{N}_m$ and $s_{m,n}^{(q)}=0$ if $ n\notin\mathcal{N}_m$, $q=1,\cdots,Q$. We further let $\boldsymbol{s}_m^{(q)}=[s_{m,1}^{(q)},\cdots,s_{m,N}^{(q)}]^{T}$ denote the signals of the $m$-th BS over all the $N$ sub-carriers in the $q$-th OFDM symbol. Assume that each BS $m$, $m=1,2$, transmits with identical power at all its assigned sub-carriers $\mathcal{N}_m$, which is denoted by $p_m$. Then, the time-domain downlink OFDM signal of the $m$-th BS in the $q$-th OFDM symbol duration is denoted by
\begin{align}\label{eq:tm_sg}
    \boldsymbol{\chi}_m^{(q)}=[\chi_{m,1}^{(q)},...,\chi_{m,N}^{(q)}]^T
    =\boldsymbol{{W}}^H\sqrt{p_m}\boldsymbol{{s}}_{m}^{(q)},\quad\forall m, q,
\end{align}where $\chi_{m,n}^{(q)}$ denotes the $n$-th sample transmitted by the $m$-th
BS in the $q$-th OFDM symbol duration, and $\boldsymbol{{W}}\in \mathbb{C}^{N\times N}$ denotes the $N\times N$ discrete Fourier transform (DFT) matrix. At the beginning of each OFDM symbol $q$, a cyclic prefix (CP) consisting of $J$ OFDM samples is inserted to eliminate the inter-symbol interference. Therefore, the overall time-domain transmitted signal by the $m$-th BS in the $q$-th OFDM symbol duration is expressed as
\begin{align}
\boldsymbol{\bar{\chi}}_m^{(q)}=[\underbrace{\chi_{m, N-J+1}^{(q)}, \ldots, \chi_{m, N}^{(q)}}_{\mathrm{CP}}, \underbrace{\chi_{m, 1}^{(q)}, \ldots, \chi_{m, N}^{(q)}}_{\text {pilot or data }}]^T, \quad\forall m.
\end{align}

Similar to the works focusing on the classic trilateration-based localization theory \cite{dvc,Torrieri84,Mao07}, we assume that there only exist line-of-sight (LOS) paths between the two BSs, between each BS and each IRS, between each BS and each target, and between each target and its associated IRS. Thus, after emitting the OFDM signals $\boldsymbol{\bar{\chi}}_m^{(q)}$'s, $q=1,\cdots,Q$, BS $m$ can receive echoes from the following four types of links.

\begin{itemize}
\item[1.] \emph{Type I Link}: the LOS link from BS $\bar{m}$ to BS $m$, where $\bar{m}=2$ if $m=1$, and $\bar{m}=1$ otherwise.
%\begin{align}
%\bar{m}=\left\{\begin{array}{ll} 1, & {\rm if} ~ m=2, \\ 2, & {\rm if} ~ m=1.\end{array} \right.
%\end{align}
\item[2.] \emph{Type II Link}: the LOS link from BS $u$ to IRS $r$ to BS $m$. Because $u\in \{1,2\}$ and $r\in \{1,\cdots,R\}$, there are $2R$ Type II links for each BS $m$.
\item[3.] \emph{Type III Link}: the LOS link from BS $u$ to target $k$ to BS $m$. Because $u\in \{1,2\}$ and $k\in \{1,\cdots,K\}$, there are $2K$ Type III links for each BS $m$.
\item[4.] \emph{Type IV Link}: the LOS link from BS $u$ to target $k$ and its associated IRS $\gamma_k$ to BS $m$. Because $u\in \{1,2\}$ and $k\in \{1,\cdots,K\}$, there are $2K$ Type IV links for each BS $m$.
\end{itemize}

For Type I links, define $h_{\bar{m},m}^{\text{I}}$ as the channel from BS $\bar{m}$ to BS $m$, $\forall m$. For Type II links, define $h_{u,m,r}^{\text{II},(q)}$ as the cascaded channel from BS $u$ to IRS $r$ to BS $m$ at the $q$-th OFDM symbol, which depends on the IRS reflecting elements at the $q$-th OFDM symbol, $\forall u,m,r,q$. For Type III links, define $h_{u,m,k}^{\text{III}}$ as the channel from BS $u$ to target $k$ to BS $m$, $\forall u,m,k$. Last, for Type IV links, define $h_{u,m,k,\gamma_k}^{\text{IV},(q)}$ as the cascaded channel from BS $u$ to target $k$ and its associated IRS $\gamma_k$ to BS $m$ at the $q$-th OFDM symbol, which depends on the IRS reflecting elements at the $q$-th OFDM symbol, $\forall u, m, k, q$. Thus, at the $n$-th sample period of the $q$-th OFDM symbol duration, the received signal at the $m$-th BS's radar receive antenna is given as
\begin{align}\label{eq:rcvd_sig_smpl}
    y_{m,n}^{(q)}&=\underbrace{h_{\bar{m},m}^{\text{I}}\bar\chi_{\bar{m},n-l_{\bar{m},m}^{\text{I}}}^{(q)}}_{\emph{Type I Link}}+\underbrace{\sum_{u=1}^{2}\sum_{r=1}^{R}h_{u,m,r}^{\text{II},(q)}\bar{\chi}_{u,n-l_{u,m,r}^{\text{II}}}^{(q)}}_{\emph{Type II Link}}\notag\\&+\underbrace{\sum_{u=1}^{2}\sum_{k=1}^{K} h_{u,m,k}^{\text{III}}\bar{\chi}_{u,n-l_{u,m,k}^{\text{III}}}^{(q)}}_{\emph{Type III Link}}+\underbrace{\sum_{u=1}^{2}\sum_{k=1}^{K} h_{u,m,k,\gamma_k}^{\text{IV},(q)}\bar{\chi}^{(q)}_{u,n-l_{u,m,k,\gamma_k}^{\text{IV}}}}_{\emph{Type IV Link}}+z_{m,n}^{(q)}, \quad \forall m,n,q,
\end{align}where $l_{\bar{m},m}^{\text{I}}$, $l_{u,m,r}^{\text{II}}$, $l_{u,m,k}^{\text{III}}$, and $l_{u,m,k,\gamma_k}^{\text{IV}}$ denote the propagation delays (in terms of OFDM samples) from BS $\bar{m}$ to BS $m$, from BS $u$ to IRS $r$ to BS $m$, from BS $u$ to target $k$ to BS $m$, and from BS $u$ to target $k$ and its associated IRS $\gamma_k$ to BS $m$, respectively, and $z_{m,n}^{(q)}\sim \mathcal{CN}(0,(\sigma_{m}^{(q)})^2)$ denotes the Gaussian noise of BS $m$ at the $n$-th sample period of the $q$-th OFDM symbol duration.

For convenience, define
\begin{align}
\boldsymbol{h}_{u,m}^{(q)}=[h_{u,m,0}^{(q)},h_{u,m,1}^{(q)},\cdots,h_{u,m,L-1}^{(q)}]^T, ~ \forall u, m, q,
\end{align}as the $L$-tap multi-path channel from BS $u$ to BS $m$ at the $q$-th OFDM symbol duration, where $L$ denotes the number of detectable paths, and the channel with a delay of $l$ OFDM samples is given as
\begin{align}\label{eqn:multi-path channel}
h_{u,m,l}^{(q)}=\left\{\begin{array}{ll} h_{\bar{m},m}^{\text{I}}, & {\rm if} ~ l=l_{\bar{m},m}^{\text{I}} ~ {\rm and} ~ u=\bar{m}, \\ h_{u,m,r}^{\text{II},(q)}, & {\rm if} ~ l=l_{u,m,r}^{\text{II}}, \\  h_{u,m,k}^{\text{III}}, & {\rm if} ~ l=l_{u,m,k}^{\text{III}}, \\ h_{u,m,k,\gamma_k}^{\text{IV},(q)}, & {\rm if} ~ l=l_{u,m,k,\gamma_k}^{\text{IV}}, \\ 0, & {\rm otherwise} .\end{array} \right.
\end{align}Then, the time-domain received signal model (\ref{eq:rcvd_sig_smpl}) is equivalent to
\begin{align}\label{eqn:signal}
y_{m,n}^{(q)}=\sum\limits_{u=1}^2\sum\limits_{l=0}^{L-1}h_{u,m,l}^{(q)}\bar\chi_{u,n-l}^{(q)}+z_{m,n}^{(q)}, \quad \forall m,n,q.
\end{align}

After removing the CP and performing the DFT operation, the received signal of BS $m$ over its assigned sub-carriers of the $q$-th OFDM symbol, i.e., $\mathcal{N}_m$, is given as
\begin{align}\label{eq:rcvd_sig}
    {\boldsymbol{\tilde{y}}}_{m}^{(q)}&=[{\tilde{y}}_{m,\mathcal{N}_m(1)}^{(q)},\cdots,{\tilde{y}}_{m,\mathcal{N}_m(|\mathcal{N}_m|)}^{(q)}]^T=\!\sqrt{p_m}\text{diag}(\boldsymbol{\tilde{s}}_m^{(q)})\boldsymbol{G}_m\boldsymbol{h}_{m,m}^{(q)}\!
    +\!\boldsymbol{\tilde{z}}_m^{(q)},\quad\forall m,q,
\end{align}where $\boldsymbol{\tilde{s}}_m^{(q)}=[s_{m,\mathcal{N}_m(1)},\cdots,s_{m,\mathcal{N}_m(|\mathcal{N}_m|)}]^T$ is the collection of signals transmitted by BS $m$ at its assigned sub-carriers of the $q$-th OFDM symbol, $\boldsymbol{G}_m\in \mathbb{C}^{|\mathcal{N}_m|\times L}$ with the element on the $n$-th row and $l$-th column denoted by $G_{m,n,l}=e^ {(\frac{-j 2 \pi(\mathcal{N}_m(n)-1)l}{N})}$, and $\boldsymbol{\tilde{z}}_m^{(q)}\sim \mathcal{CN}(0,(\sigma_{m}^{(q)})^2 \boldsymbol{I})$ is the noise at the $m$-th BS over its assigned sub-carriers of the $q$-th OFDM symbol. Note that because BS $\bar{m}$ transmits at its own sub-carriers $\mathcal{N}_{{\bar{m}}}$ in the frequency domain, it does not contribute to the received signals of BS $m$ at sub-carriers $\mathcal{N}_m$.

\vspace{-5mm}
\section{A Two-Phase Localization Protocol}\label{sec:protocol}
The key observation from (\ref{eqn:multi-path channel}) is that rich delay information is embedded in the channels $\boldsymbol{h}_{m,m}^{(q)}$'s, $m=1,2$, which can be estimated via (\ref{eq:rcvd_sig}). Specifically, if $h_{m,m,l}^{(q)}\neq 0$ for some $l$, it then indicates the existence of a path from BS $m$ to some IRS/target back to BS $m$ with a delay of $l$ OFDM samples. This path may arise from a Type II link such that the propagation delay from BS $m$ to some IRS $r$ to BS $m$ is $l_{m,m,r}^{\text{II}}=l$; or a Type III link such that the propagation delay from BS $m$ to some target $k$ to BS $m$ is $l_{m,m,k}^{\text{III}}=l$; or a Type IV link such that the propagation delay from BS $m$ to some target $k$ and its associated IRS $\gamma_k$ to BS $m$ is $l_{m,m,k,\gamma_k}^{\text{IV}}=l$. Based on this observation, we propose a two-phase localization protocol in this paper. In the first phase (Phase I), each BS $m$ estimates the channels $\boldsymbol{h}_{m,m}^{(q)}$'s based on the echoes (\ref{eq:rcvd_sig}) and extracts the delay information (thus the range information) from the estimated channels. After Phase I, BS $m$ can obtain two sets that contain the range information of the targets, i.e., $\mathcal{D}_m^{\text{III}}=\{\hat{d}_{m,1}^{\text{III}},\cdots,\hat{d}_{m,K}^{\text{III}}\}$, where $\hat{d}_{m,k}^{\text{III}}$ denotes the estimated range of a Type III link from BS $m$ to target $k$ to BS $m$, $\forall k$, and $\mathcal{D}_m^{\text{IV}}=\{\hat{d}_{m,1,\gamma_1}^{\text{IV}},\cdots,\hat{d}_{m,K,\gamma_K}^{\text{IV}}\}$, where $\hat{d}_{m,k,\gamma_k}^{\text{IV}}$ denotes the estimated range of a Type IV link from BS $m$ to target $k$ and its associated IRS $\gamma_k$ to BS $m$, $\forall k$, $m=1,2$. Then, in the second phase (Phase II), two BSs send their range estimation sets obtained in Phase I to the central processor, which then localizes the targets using the trilateration method. The feasibility of using BS $1$, BS $2$, and IRS $\gamma_k$ as three anchors to localize target $k$ is as follows. First, the distance from target $k$ to BS $m$, i.e., $d_{m,k}^{\text{BT}}$, can be estimated from $\hat{d}_{m,k}^{\text{III}}$, $m=1,2$. Second, although the distance between target $k$ and its associated IRS $\gamma_k$ cannot be directly estimated, it can be indirectly estimated from the Type III and Type IV links based on the following equation
\begin{align}\label{eq:d_it_equal}
    d_{\gamma_k,k}^{\text{IT}}=d_{1,k,\gamma_k}^{\text{IV}}-d_{1,k}^{\text{BT}}-d_{1,\gamma_k}^{\text{BI}}=d_{2,k,\gamma_k}^{\text{IV}}-d_{2,k}^{\text{BT}}-d_{2,\gamma_k}^{\text{BI}},\quad \forall k,
\end{align}where $d_{m,k,\gamma_k}^{\text{IV}}$ denotes the true range of the Type IV link from BS $m$ to target $k$ and its associated IRS $\gamma_k$ to BS $m$. In (\ref{eq:d_it_equal}), $d_{m,\gamma_k}^{\text{BI}}$ is known given the locations of the BSs and the IRSs, and  $d_{m,k,\gamma_k}^{\text{IV}}$ can be estimated in Phase I as $\hat{d}_{m,k,\gamma_k}^{\text{IV}}$. With the distances from target $k$ to BS $1$, BS $2$, and IRS $\gamma_k$, we can thus efficiently localize this target.

However, we encounter the so-called data association issue in Phase II \cite{dvc}, which is common in device-free sensing \cite{Torrieri84}. In our considered IRS-assisted sensing system, there are two challenges arising from data association. First, although the estimated ranges of Type III and Type IV links are available, the central processor does not know how to match each element in $\mathcal{D}_m^{\text{III}}$ and $\mathcal{D}_m^{\text{IV}}$ to the right target, $m=1,2$. Second, when estimating the distance between target $k$ and its associated IRS $\gamma_k$ based on (\ref{eq:d_it_equal}), besides the mapping of the right elements in $\mathcal{D}_m^{\text{III}}$ and $\mathcal{D}_m^{\text{IV}}$ to target $k$, we do not know the associated IRS to target $k$ neither, because before localizing target $k$, we do not know which IRS is closest to it. Without the knowledge of $\gamma_k$, a wrong distance between BS $m$ and some other IRS may be deducted when we calculate $d_{\gamma_k,k}^{\text{IT}}$ based on (\ref{eq:d_it_equal}). Moreover, after $d_{\gamma_k,k}^{\text{IT}}$ is estimated, we may use a wrong IRS $r \neq \gamma_k$ as the anchor to localize target $k$.

In the rest of this paper, we introduce how to estimate the ranges of Type III and Type IV links based on (\ref{eq:rcvd_sig}) in Phase I, and how to tackle the data association issue when localizing the targets based on the trilateration method in Phase II.

\vspace{-5mm}
\section{Phase I: Range Estimation Based on IRS ON-OFF Scheme}\label{sec:range_est}
As explained in Section \ref{sec:protocol}, the goal of Phase I of our proposed protocol is to first estimate the multi-path channels $\boldsymbol{h}_{m,m}^{(q)}$'s based on the received signal \eqref{eq:rcvd_sig} and then estimate the delay/range of Type III and Type IV links based on \eqref{eqn:multi-path channel}. In this paper, we adopt an on-off scheme to control the IRSs over $Q=2$ OFDM symbols so as to estimate the Type III channels during IRS's ``off'' state and Type IV channels during IRS's ``on'' state separately. Under this scheme, over OFDM symbol $q=1$, all IRSs are in the absorbing mode (the mobile users perform downlink communication without the assist of the IRS in this symbol duration) such that the frequency-domain echoes received by the BSs given in (\ref{eq:rcvd_sig}) are merely contributed by the Type III links; while over OFDM symbol $q=2$, all IRSs are in the reflecting mode such that the received echoes are contributed by Type II, Type III, and Type IV links.

Specifically, when $q=1$, because all IRSs are in the absorbing mode, we have $h_{m,m,r}^{\text{II},(1)}=0$ and $h_{m,m,k,\gamma_k}^{\text{IV},(1)}=0$, $\forall m, k, r$. In this case, the multi-path channel \eqref{eqn:multi-path channel} reduces to
\begin{align}\label{eqn:channel off}
    h_{m,m,l}^{(1)}=\left\{\begin{array}{ll} h_{m,m,k}^{\text{III}}, & {\rm if} ~ l=l_{m,m,k}^{\text{III}} , \\ 0, & {\rm otherwise} ,\end{array} \right. \quad m=1,2.
\end{align}Because $\boldsymbol{h}_{m,m}^{(1)}$'s are sparse channel vectors according to (\ref{eqn:channel off}), we can use the LASSO technique to estimate them based on \eqref{eq:rcvd_sig}, by solving the following problem
\begin{align} \label{eqn:problem 1}
\mathop{\mathrm{min}}_{\boldsymbol{h}_{m,m}^{(1)}}    0.5\left\|\boldsymbol{\tilde{y}}_m^{(1)}-\sqrt{p_m}\text{diag}(\tilde{\boldsymbol{s}}_m^{(1)})\boldsymbol{G}_m\boldsymbol{h}_{m,m}^{(1)} \right\|_{{2}}^2+\rho\left\|\boldsymbol{h}_{m,m}^{(1)} \right\|_1,
\end{align}where $\rho \geq 0$ is a given coefficient to control the sparsity of the estimated channels. Let $\hat{\boldsymbol{h}}_{m,m}^{(1)}=[\hat{h}_{m,m,0}^{(1)},\ldots,\hat{h}_{m,m,L-1}^{(1)}]^T$ denote the optimal solution to the above convex problem, $m=1,2$. Because $\hat{\boldsymbol{h}}_{m,m}^{(1)}$'s are imperfect estimations, we estimate the support of $\boldsymbol{h}_{m,m}^{(1)}$'s based on a threshold-based strategy. Specifically, we declare $h_{m,m,l}^{(1)}\neq 0$ if and only if $\|\hat{h}_{m,m,l}^{(1)}\|_2\geq \delta_1$, $\forall m,l$, where $\delta_1$ is some given threshold. Then, given any $l \in \Phi_m^{\text{III}}=\{l|\|\hat{h}_{m,m,l}^{(1)}\|_2\geq \delta_1\}$, we declare that there exists a target $\bar{k}_{m,l}$ such that the range of a Type III link from BS $m$ to target $\bar{k}_{m,l}$ to BS $m$, i.e., $d_{m,\bar{k}_{m,l}}^{\text{III}}$, can be estimated as \cite{dvc}
\begin{align}\label{eq:range 1}
\hat{d}_{m,\bar{k}_{m,l}}^{\text{III}}=\frac{lc_0}{N\Delta f}+\frac{c_0}{2N\Delta f},\quad  m=1,2,
\end{align}where $c_0$ denotes the speed of the light. To summarize, after the first OFDM symbol duration, each BS $m$ will have a set consisting of the estimated ranges of Type III links, i.e.,
\begin{align}\label{eqn:range set 1}
\mathcal{D}_m^{\text{III}}=\{\hat{d}_{m,\bar{k}_{m,l}}^{\text{III}}|\forall l \in \Phi_m^{\text{III}}\}, \quad  m=1,2.
\end{align}

Over the period of OFDM symbol $q=2$, all the IRSs are in the reflecting mode. Therefore, the received signals ${\boldsymbol{\tilde{y}}}_{m}^{(2)}$'s given in (\ref{eq:rcvd_sig}) are contributed by Type II, Type III, and Type IV links. As shown in (\ref{eqn:multi-path channel}), the channels to be estimated, i.e., $\boldsymbol{h}_{m,m}^{(2)}$'s, $m=1,2$, are sparse. Therefore, similar to the estimation of $\boldsymbol{h}_{m,m}^{(1)}$'s, we can apply the LASSO technique to estimate $\boldsymbol{h}_{m,m}^{(2)}$'s based on the received signals ${\boldsymbol{\tilde{y}}}_{m}^{(2)}$'s. However, when $q=2$, we have some side information about the support of $\boldsymbol{h}_{m,m}^{(2)}$'s. First, because we know the locations of the BSs and the IRSs, the delay of Type II links is known. Define
$\Phi_m^{\text{II}}=\{l|\frac{lc_0}{N\Delta f} \leq d_{m,r}^{\text{BI}}\leq \frac{(l+1)c_0}{N\Delta f}, ~ r=1,\cdots,R\}$ as the collection of the delay (in terms of OFDM samples) of Type II links to BS $m$, $m=1,2$. Then, we have $h_{m,m,l}^{(2)}\neq 0$ if $l\in \Phi_m^{\text{II}}$. Second, for Type III links, which have nothing to do with the on/off state of the IRS and possess the same delay when $q=1$ and $q=2$, we have $h_{m,m,l}^{(2)}\neq 0$ if $l\in \Phi_m^{\text{III}}$. With the above side information about the support of $\boldsymbol{h}_{m,m}^{(2)}$'s, we apply the weighted-LASSO technique to estimate $\boldsymbol{h}_{m,m}^{(2)}$, by solving the following problem \cite{weighted_lasso,weighted_lasso_3,weighted_lasso_4}
\begin{align}
    \mathop{\mathrm{min}}_{\boldsymbol{h}_{m,m}^{(2)}}   0.5\left\|\boldsymbol{\tilde{y}}_m^{(2)}-\sqrt{p_m}\text{diag}(\tilde{\boldsymbol{s}}_m^{(2)})\boldsymbol{G}_m\boldsymbol{h}_{m,m}^{(2)}\right\|_{2}^2+\sum_{l=0}^{L-1}\beta_l\left\|h_{m,m,l}^{(2)} \right\|_1,
\end{align}where \begin{align}
   \beta_l= \left\{\begin{array}{ll}
 \rho_1, & \text{if}~l\in \Phi_{m}^{\text{II}}\cup \Phi_{m}^{\text{III}},\\
 \rho_2, & \text{otherwise},~~~~~~
\end{array}\right.\quad\forall l,
\end{align}is a binary threshold with $\rho_1<\rho_2$. Let $\hat{\boldsymbol{h}}_{m,m}^{(2)}=[\hat{h}_{m,m,0}^{(2)},\ldots,\hat{h}_{m,m,L-1}^{(2)}]^T$ denote the optimal solution to the above convex problem, $m=1,2$. Similar to Phase I, given some threshold $\delta_2$, define $\Phi_m=\{l|\|{\hat{h}}_{m,m,l}^{(2)}\|_2\geq  \delta_2\}$, $m=1,2$. Then, if $l\in \Phi_m$, we declare that $h_{m,m,l}^{(2)}\neq 0$. The purpose of using the binary threshold is to force all the elements in $\Phi_{m}^{\text{II}}\cup \Phi_{m}^{\text{III}}$ to be contained in the set $\Phi_m$, i.e., Type II and Type III links can be detected via the weighted-LASSO technique. Define $\Phi_m^{\text{IV}}=\{l|l\in \Phi_m ~ \text{but} ~ l \notin \Phi_m^{\text{II}} \cup \Phi_m^{\text{III}}\}$ as the set containing all the estimated delay (in terms of OFDM samples) of Type IV links to BS $m$, $m=1,2$. As a result, given any $l\in \Phi_m^{\text{IV}}$, we declare that there exists a target $\tilde{k}_{m,l}$ such that the range of a Type IV link from BS $m$ to target $\tilde{k}_{m,l}$ and its associated IRS $\gamma_{\tilde{k}_{m,l}}$ to BS $m$, i.e., $d_{m,\tilde{k}_{m,l},\gamma_{\tilde{k}_{m,l}}}^{\text{IV}}$, can be estimated as \cite{dvc}
\begin{align}\label{eq:range 2}
\hat{d}_{m,\tilde{k}_{m,l},\gamma_{\tilde{k}_{m,l}}}^{\text{IV}}=\frac{lc_0}{N\Delta f}+\frac{c_0}{2N\Delta f},\quad m=1,2.
\end{align}To summarize, after the second OFDM symbol duration, each BS $m$ will have another distance set consisting of the estimated ranges of Type IV links, i.e.,
\begin{align}\label{eqn:distance set 2}
\mathcal{D}_m^{{\text{IV}}}=\{\hat{d}_{m,\tilde{k}_{m,l},\gamma_{\tilde{k}_{m,l}}}^{\text{IV}}| \forall l \in \Phi_m^{\text{IV}}\}, \quad m=1,2.
\end{align}
\section{Phase II: Data Association and Localization}\label{sec:da_loc}
In Phase II, we need to localize the $K$ targets based on the knowledge about $\mathcal{D}_m^{{\text{III}}}$'s and $\mathcal{D}_m^{{\text{IV}}}$'s obtained in Phase I. As discussed in Section \ref{sec:protocol}, the main challenge for localization lies in data association, i.e., $\bar{k}_{m,l}$'s, $\tilde{k}_{m,l}$'s, and $\gamma_k$'s are unknown in $\mathcal{D}_m^{{\text{III}}}$'s and $\mathcal{D}_m^{{\text{IV}}}$'s. Without such knowledge, when we aim to use the ranges from target $k$ to the three anchors, i.e., BS $1$, BS $2$, and IRS $\gamma_k$, for localizing target $k$ based on the trilateration method, we do not know which elements in $\mathcal{D}_m^{{\text{III}}}$'s and $\mathcal{D}_m^{{\text{IV}}}$'s are target $k$'s ranges, and which IRS is the anchor for target $k$. In this section, we tackle the data association issue such that the trilateration method can be applied in our considered IRS-assisted networked sensing architecture.
\vspace{-5mm}
\subsection{General Framework for Data Association and Localization}

For convenience, define $\lambda_{m,k}\in\{1,\cdots,K\}$ such that the estimated range for the path from BS $m$ to target $k$ to BS $m$, i.e., $\hat{d}_{m,k}^{\text{III}}$ shown in \eqref{eq:range 1}, is the $\lambda_{m,k}$-th largest element in $\mathcal{D}_m^{\text{III}}$. Moreover, define $\mu_{m,k}\in\{1,\cdots,K\}$ such that the estimated range of the path from BS $m$ to target $k$ and its associated IRS $\gamma_k$ to BS $m$, i.e., $\hat{d}_{m,k,\gamma_k}^{\text{IV}}$ shown in \eqref{eq:range 2}, is the $\mu_{m,k}$-th largest element in $\mathcal{D}_{m}^{\text{IV}}$. In other words, we have $\hat{d}_{m,k}^{\text{III}}=\mathcal{D}_{m}^{\text{III}}(\lambda_{m,k})$ and $\hat{d}_{m,k,\gamma_k}^{\text{IV}}=\mathcal{D}_{m}^{\text{IV}}(\mu_{m,k})$, $\forall m,k$, where given any set $\mathcal{A}$, $\mathcal{A}(a)$ denotes the $a$-th largest element in $\mathcal{A}$. Given the definitions of $\lambda_{m,k}$'s and $\mu_{m,k}$'s, according to \eqref{eq:d_it_equal}, the distance between target $k$ and its associated IRS $\gamma_k$ estimated from the range information obtained by BS $m$ is expressed as
\begin{align}\label{eq:est_d_it}
    \hat{d}_{\gamma_k,k}^{\text{IT},m}=\hat{d}_{m,k,\gamma_k}^{\text{IV}}-\hat{d}_{m,k}^{\text{BT}}-d_{m,\gamma_k}^{\text{BI}}=\mathcal{D}_m^{\text{IV}}(\mu_{m,k})-\frac{1}{2}\mathcal{D}_m^{\text{III}}(\lambda_{m,k})-d_{m,\gamma_k}^{\text{BI}},\quad m=1, 2, ~ \forall k,
\end{align}where
\begin{align}\label{eq:est_IRS}
\hat{d}_{m,k}^{\text{BT}}=\mathcal{D}_m^{\text{III}}(\lambda_{m,k})/2, \quad \forall m,k,
\end{align}is the estimation of the distance between target $k$ and BS $m$, i.e., $d_{m,k}^{\text{BT}}$. Therefore, there are two estimations of $d_{\gamma_k,k}^{\text{IT}}$, i.e., $\hat{d}_{\gamma_k,k}^{\text{IT},1}$ obtained by the range information from BS $1$, i.e., $\mathcal{D}_1^{{\text{III}}}$ and $\mathcal{D}_1^{{\text{IV}}}$, and $\hat{d}_{\gamma_k,k}^{\text{IT},2}$ obtained by the range information from BS $2$, i.e., $\mathcal{D}_2^{{\text{III}}}$ and $\mathcal{D}_2^{{\text{IV}}}$.

Note that if $\lambda_{m,k}$'s and $\mu_{m,k}$'s are known, $\bar{k}_{m,l}$'s and $\tilde{k}_{m,l}$'s are known. Therefore, we can define the set consisting of all the data association variables as $\mathcal{X}_1=\{[\lambda_{1,k},\lambda_{2,k},\mu_{1,k},\mu_{2,k},\gamma_k]|\forall k\}$. Furthermore, define $\mathcal{X}_2=\{(x_k^{\text{T}},x_k^{\text{T}})|\forall k\}$ as the set consisting of all the target location variables. Then, we aim to estimate the data association and target location variables, i.e., $\mathcal{X}_1$ and $\mathcal{X}_2$, based on the following equations
\begin{align}
    &\frac{1}{2}\mathcal{D}_m^{\text{III}}(\lambda_{m,k})=\sqrt{(x_m^{\text{B}}-x_k^{\text{T}})^2+(y_m^{\text{B}}-y_k^{\text{T}})^2}+\epsilon_{m,k},\quad \forall m,k,\label{eq:d_bt_noise}\\
    &\mathcal{D}_m^{\text{IV}}(\mu_{m,k})-\frac{1}{2}\mathcal{D}_m^{\text{III}}(\lambda_{m,k})-d_{m,\gamma_k}^{\text{BI}}=\sqrt{(x_{\gamma_k}^{\text{I}}-x_k^{\text{T}})^2+(y_{\gamma_k}^{\text{I}}-y_k^{\text{T}})^2}+\varsigma_{m,k},\quad \forall m,k,\label{eq:d_bitb_noise}\\
    &\{\lambda_{m,1},\cdots,\lambda_{m,K}\}=\{1,\cdots,K\},
    \quad \forall m,\label{eq:con_lambda}\\
    &\{\mu_{m,1},\cdots,\mu_{m,K}\}=\{1,\cdots,K\},\quad \forall m,\label{eq:con_mu}\\
    & \gamma_k\in \{1,\cdots,R\}, \quad \forall k, \label{eq:con gamma}
\end{align}where (\ref{eq:con_lambda}) and (\ref{eq:con_mu}) guarantee that different elements in $\mathcal{D}_m^{{\text{III}}}$'s and $\mathcal{D}_m^{{\text{IV}}}$'s are assigned to different targets, (\ref{eq:con gamma}) guarantees that each target is associated with one of the $R$ IRSs, and $\epsilon_{m,k}$ and $\varsigma_{m,k}$ denote the unknown errors for estimating $d_{m,k}^{\text{BT}}$ and $d_{\gamma_k,k}^{\text{IT}}$ as $\hat{d}_{m,k}^{\text{BT}}$ given in (\ref{eq:est_IRS}) and $\hat{d}_{\gamma_k,k}^{\text{IT}}$ given in (\ref{eq:est_d_it}), respectively. In the literature of localization, it is usually assumed that the range estimation errors are Gaussian errors \cite{Torrieri84, Mao07}, i.e., $\epsilon_{m,k} \sim \mathcal{CN}(0,\hat{\sigma}_{m,k}^{2})$ and $\varsigma_{m,k} \sim \mathcal{CN}(0,\tilde{\sigma}_{m,k}^{2})$, $\forall m,k$.

One straightforward method to jointly estimate $\mathcal{X}_1$ and $\mathcal{X}_2$ is to perform exhaustive search over all the feasible data association solutions satisfying \eqref{eq:con_lambda}, \eqref{eq:con_mu}, and \eqref{eq:con gamma} for finding the one that yields the best localization solution. In particular, define
\begin{align}
\bar{\mathcal{Y}}=\{\mathcal{X}_1|(\ref{eq:con_lambda}), ~ (\ref{eq:con_mu}), ~ \text{and} ~ (\ref{eq:con gamma}) ~ \text{hold}\},
\end{align}as the set consisting of all the data association solutions satisfying \eqref{eq:con_lambda}, \eqref{eq:con_mu}, and \eqref{eq:con gamma}. Given any feasible data association solution $\mathcal{X}_1\in \bar{\mathcal{Y}}$, the corresponding localization solution of target $k$ can be obtained by solving the following maximum likelihood (ML) problem to (\ref{eq:d_bt_noise}) and (\ref{eq:d_bitb_noise}) \cite{Torrieri84, Mao07}
\begin{align}\label{prb:loc_with_da}
&\underset{(x_k^{\text{T}},y_k^{\text{T}})}{\text{minimize}}~\sum_{m=1}^{2} (\hat{f}_{m,k}(\lambda_{m,k},x_k^{\text{T}},y_k^{\text{T}})+\tilde{f}_{m,k}(\lambda_{m,k},\mu_{m,k},\gamma_k,x_k^{\text{T}},y_k^{\text{T}})),
\end{align}where
\begin{align}
    &\hat{f}_{m,k}(\lambda_{m,k},x_k^{\text{T}},y_k^{\text{T}})=
    \frac{\!(\frac{1}{2}\mathcal{D}_{m}^{\text{III}}(\lambda_{m, k})-\sqrt{(x_m^{\text{B}}-x_k^{\text{T}})^2+(y_m^{\text{B}}-y_k^{\text{T}})^2})^{2}}{{ \hat{\sigma}_{m, k}^2}},\quad \forall m,k, \label{eq:cost_d_bi}\\
    &\tilde{f}_{m,k}(\lambda_{m,k},\mu_{m,k},\gamma_k,x_k^{\text{T}},y_k^{\text{T}})\nonumber \\ =&\frac{(\mathcal{D}_{m}^{\text{IV}}(\mu_{m, k})-\frac{1}{2}\mathcal{D}_{m}^{\text{III}}(\lambda_{m, k})-d_{m,\gamma_k}^{\text{BI}}-\sqrt{(x_{\gamma_k}^{\text{I}}-x_k^{\text{T}})^2+(y_{\gamma_k}^{\text{I}}-y_k^{\text{T}})^2})^{2}}{{\tilde{\sigma}_{m, k}^2}},\quad \forall m,k,\label{eq:cost_d_bitb}
\end{align}denote the estimation residuals for target $k$. Similar to \cite{Torrieri84,Mao07}, we can apply the Gauss-Newton method to solve the above non-linear non-convex problem. Let $(x_k^{\text{T}}(\mathcal{X}_{1,k}),y_k^{\text{T}}(\mathcal{X}_{1,k}))$ denote the location solution of target $k$ corresponding to the given data association solution, where $\mathcal{X}_{1,k}=[\lambda_{1,k},\lambda_{2,k},\mu_{1,k},\mu_{2,k},\gamma_k]$, $\forall k$, denotes the data association solution for target $k$. Then, the optimal data association solution can be obtained via exhaustive search by solving the following problem
\begin{align}\label{prb:da_with_loc}
&\underset{{\mathcal{X}_1}}{\text{minimize}}~\sum_{k=1}^{K}  \sum_{m=1}^{2} (\hat{f}_{m,k}(\lambda_{m,k},\!x_k^{\text{T}}(\mathcal{X}_{1,k})\!,\!y_k^{\text{T}}(\mathcal{X}_{1,k}))\!+\!\tilde{f}_{m,k}(\lambda_{m,k},\!\mu_{m,k},\!\gamma_k,\!x_k^{\text{T}}(\mathcal{X}_{1,k}),\!y_k^{\text{T}}(\mathcal{X}_{1,k})))\\
&\text{subject to}~\mathcal{X}_1\in \bar{\mathcal{Y}}.\notag
\end{align}The objective function of the above problem is the sum of the estimation residuals of all the targets. After the optimal data association solution $\mathcal{X}_1^{\ast}$ is obtained, the solution to problem \eqref{prb:loc_with_da} given this data association solution can be used as the final localization solution.

However, the number of feasible data association solutions in the set $\bar{\mathcal{Y}}$ that satisfy \eqref{eq:con_lambda}, \eqref{eq:con_mu}, \eqref{eq:con gamma} is large, and it is of prohibitive complexity to solve the complicated problem \eqref{prb:loc_with_da} for all these feasible data association solutions, as required by problem \eqref{prb:da_with_loc}. To tackle the above challenge, in this paper, we reveal a hidden property of the optimal data association solution to problem (\ref{prb:da_with_loc}), which is embedded in (\ref{eq:d_it_equal}) and (\ref{eq:est_d_it}) but not fully utilized in the above scheme. According to (\ref{eq:d_it_equal}) and (\ref{eq:est_d_it}), there are two methods to calculate the distance between target $k$ and its associated IRS $\gamma_k$, either from the range sets obtained by BS $1$, i.e., $\mathcal{D}_1^{{\text{III}}}$ and $\mathcal{D}_1^{{\text{IV}}}$, or those by BS $2$, i.e., $\mathcal{D}_2^{{\text{III}}}$ and $\mathcal{D}_2^{{\text{IV}}}$. If range estimation in Phase I is perfect, then these two methods will yield the same range estimation as shown in (\ref{eq:d_it_equal}). In practice, with imperfect range estimation in Phase I, the estimation obtained by the range information of BS $1$ and that obtained by the range information of BS $2$ should be different but very close to each other. Therefore, besides (\ref{eq:con_lambda}), (\ref{eq:con_mu}), and (\ref{eq:con gamma}), the data association solution should also satisfy the following condition
\begin{align}\label{eq:con new}
\left|\left(\mathcal{D}_1^{\text{IV}}(\mu_{1,k})-\frac{1}{2}\mathcal{D}_1^{\text{III}}(\lambda_{1,k})-d_{1,\gamma_k}^{\text{BI}}\right)-\left(\mathcal{D}_2^{\text{IV}}(\mu_{2,k})-\frac{1}{2}\mathcal{D}_2^{\text{III}}(\lambda_{2,k})-d_{2,\gamma_k}^{\text{BI}}\right)\right|\leq \tau, ~ \forall k,
\end{align}where $\tau\geq 0$ is some given threshold. Then, we define a new set as
\begin{align}
\mathcal{Y}= \{\mathcal{X}_1|(\ref{eq:con_lambda}), ~ (\ref{eq:con_mu}), ~ (\ref{eq:con gamma}), ~\text{and} ~ (\ref{eq:con new}) ~ \text{hold}\}.
\end{align}The new joint data association and localization problem can thus be formulated as
\begin{align}\label{prb:da_with_loc_1}
&\underset{{\mathcal{X}_1},\mathcal{X}_2}{\text{minimize}}~\sum_{m=1}^{2} \sum_{k=1}^{K} (\hat{f}_{m,k}(\lambda_{m,k},x_k^{\text{T}},y_k^{\text{T}})+\tilde{f}_{m,k}(\lambda_{m,k},\mu_{m,k},\gamma_k,x_k^{\text{T}},y_k^{\text{T}}))\\
&\text{subject to}~\mathcal{X}_1\in \mathcal{Y}.\notag
\end{align}

\begin{figure}[t]
	\centering
	\includegraphics[height=6.6cm]{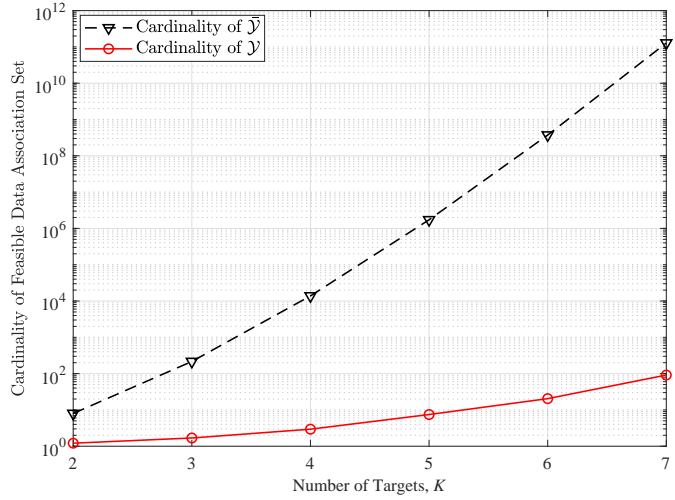}
 \vspace{-5mm}
    \caption{Comparison between the cardinality of $\bar{\mathcal{Y}}$ and that of $\mathcal{Y}$.} \label{fig:compare_dimension}
\vspace{-8mm}
\end{figure}

In the following, we provide a numerical example to verify the effectiveness of using (\ref{eq:con new}) to reduce the cardinality of $\bar{\mathcal{Y}}$. In this example, we assume that there is one IRS in the network such that $\gamma_k=1$, $\forall k$. In other words, the unknown data association variables are $\lambda_{m,k}$'s and $\mu_{m,k}$'s, and the cardinality of $\bar{\mathcal{Y}}$ is $(K!)^3$. However, the cardinality of $\mathcal{Y}$ depends on the range estimation sets $\mathcal{D}_m^{{\text{III}}}$'s and $\mathcal{D}_m^{{\text{IV}}}$'s, which determine the number of data association solutions satisfying (\ref{eq:con new}). We suppose that the locations of the two BSs and the IRS are given as $(-100,0)$, $(100,0)$, and $(0,40)$ in meter, while the $K$ targets are independently and uniformly distributed in a semi-circle whose center is the IRS and radius is $50$ meters. We generate $10^9$ realizations about target locations, and for each realization, we apply the method proposed in Section \ref{sec:range_est} to obtain the range estimation sets $\mathcal{D}_m^{{\text{III}}}$'s and $\mathcal{D}_m^{{\text{IV}}}$'s, where the channel bandwidth is set to be $B=400$ MHz. After $\mathcal{D}_m^{{\text{III}}}$'s and $\mathcal{D}_m^{{\text{IV}}}$'s are obtained, $\mathcal{Y}$ is obtained by setting $\tau=1.5$ in meter in (\ref{eq:con new}). The average cardinality of $\mathcal{Y}$ over all the realizations is recorded when $K$ ranges from $2$ to $7$. Fig. \ref{fig:compare_dimension} shows the cardinality of $\bar{\mathcal{Y}}$ without considering (\ref{eq:con new}) and the cardinality of $\mathcal{Y}$ with (\ref{eq:con new}) taken into account. It is observed that thanks to the exploitation of (\ref{eq:con new}), the cardinality of $\mathcal{Y}$ is significantly smaller than that of $\bar{\mathcal{Y}}$. Note that given each feasible data association solution, we have to solve problem (\ref{prb:loc_with_da}). Therefore, via utilizing (\ref{eq:con new}), we can significantly reduce the number of times to solve problem (\ref{prb:loc_with_da}) when solving problem (\ref{prb:da_with_loc_1}).

Excitingly, besides the advantage regarding to complexity, localization via solving problem (\ref{prb:da_with_loc_1}) is also theoretically optimal as promised by the following theorem.

\begin{theorem}\label{theorem1}
Suppose that range estimation in Phase I is perfect such that $\mathcal{D}_m^{\text{III}}=\{2d_{m,1}^{\text{BT}},\cdots,2d_{m,K}^{\text{BT}}\}$ and $\mathcal{D}_m^{\text{IV}}=\{d_{m,1}^{\text{BT}}+d_{\gamma_1,1}^{\text{IT}}+d_{m,\gamma_K}^{\text{BI}},\cdots,d_{m,K}^{\text{BT}}+d_{\gamma_K,K}^{\text{IT}}+d_{m,\gamma_K}^{\text{BI}}\}$, $\forall m$. Moreover, suppose that the fixed locations of any two IRSs $\gamma1$ and $\gamma2$ satisfy $d_{1,\gamma1}^{\text{BI}}-d_{2,\gamma1}^{\text{BI}}\neq d_{1,\gamma2}^{\text{BI}}-d_{2,\gamma2}^{\text{BI}}$, which indicates that the network topology satisfies the following two conditions: C1. any two IRSs are not deployed on the perpendicular line of the line connecting the two BSs; C2. any two IRSs are not deployed on a branch of a hyperbole whose foci are the two BSs. Then, if the targets are independently and uniformly distributed in the network when calculating $d_{m,k}^{\text{BT}}$'s, $d_{\gamma_k,k}^{\text{BI}}$'s, and $d_{m,\gamma_k}^{\text{BI}}$'s via \eqref{eqn:dis1}, \eqref{eqn:dis2}, and \eqref{eqn:dis3}, there almost surely exists a unique data association solution $\mathcal{X}_1$ satisfying \eqref{eq:con_lambda}, \eqref{eq:con_mu}, \eqref{eq:con gamma}, and \eqref{eq:con new} when $\tau$ is set to be $0$. Moreover, the localization solution $\mathcal{X}_2$ corresponding to this unique data association solution indicates the true locations of all the targets.
\end{theorem}

\begin{IEEEproof}
Please refer to Appendix \ref{appendix1}.
\end{IEEEproof}

Theorem \ref{theorem1} indicates that in the ideal case with perfect range estimation in Phase I, the optimal solution to problem (\ref{prb:da_with_loc_1}) is the correct solution almost surely. This provides a theoretical justification for our proposed strategy to localize the targets via problem (\ref{prb:da_with_loc_1}). In the rest of this section, we aim to solve problem (\ref{prb:da_with_loc_1}) efficiently in the practical case with imperfect range estimation in Phase II. Specifically, we will start with the special case when $R=1$ IRS is in the network such that all the targets are associated with this IRS, i.e., $\gamma_k=1$, $\forall k$. In this case, we merely need to estimate the data association variables $\lambda_{m,k}$'s and $\mu_{m,k}$'s. Then, based on the results for the above special case, we will propose an efficient algorithm to localize the targets in the general case when $R>1$ IRSs are in the network, where the data association variables $\gamma_k$'s should also be estimated.

\vspace{-5mm}
\subsection{Data Association and Localization with One IRS}\label{sec:da_loc_single}
First, we focus on the case with one IRS, where $\gamma_k=1$, $\forall k$, and the unknown data association variables are $\lambda_{m,k}$'s and $\mu_{m,k}$'s. In this case, given the range estimation sets $\mathcal{D}_m^{{\text{III}}}$'s and $\mathcal{D}_m^{{\text{IV}}}$'s obtained in Phase I, we can find all the data association solutions that satisfy (\ref{eq:con_lambda}), (\ref{eq:con_mu}), and (\ref{eq:con new}) to get $\mathcal{Y}$ (note that (\ref{eq:con gamma}) is not considered here because we know $\gamma_k=1$, $\forall k$). In the ideal case where range estimation in Phase I is perfect, Theorem \ref{theorem1} shows that there is only one unique data association solution in $\mathcal{Y}$. In the practical case with imperfect range estimation in Phase I, however, there are multiple data association solutions in $\mathcal{Y}$. Because the cardinality of $\mathcal{Y}$ is small, as verified in Fig. \ref{fig:compare_dimension}, we may solve problem (\ref{prb:da_with_loc_1}) via exhaustive search. Specifically, given any data association solution $\mathcal{X}_1$ in the set $\mathcal{Y}$, we can apply the Gauss-Newton method to solve problem (\ref{prb:loc_with_da}) and find the localization solution corresponding to this data association solution, which is denoted by $(x_k^{\text{T}}(\mathcal{X}_{1,k}),y_k^{\text{T}}(\mathcal{X}_{1,k}))$, $\forall k$. After searching all the data association solutions in $\mathcal{Y}$, the one leading to the minimum total estimation residual for all the targets will be determined to be the optimal data association solution, and the solution to problem (\ref{prb:loc_with_da}) corresponding to the optimal data association solution will be determined to be the locations of the targets.

In practice, we can further reduce the complexity of the above exhaustive search algorithm. Specifically, given any data association solution $\mathcal{X}_1\in \mathcal{Y}$, suppose that there exists a target $k$ whose location estimation satisfies
\begin{align}\label{eq:con_cost}
    \sum_{m=1}^{2} (\hat{f}_{m,k}(\lambda_{m,k},x_k^{\text{T}}(\mathcal{X}_{1,k}),y_k^{\text{T}}(\mathcal{X}_{1,k})\!+\!\tilde{f}_{m,k}(\lambda_{m,k},\mu_{m,k},\gamma_k,x_k^{\text{T}}(\mathcal{X}_{1,k}),y_k^{\text{T}}(\mathcal{X}_{1,k})))\!\geq \! \xi, ~ \forall k,
\end{align}where $\xi>0$ is some given threshold. Then, it indicates that with data association solution $\mathcal{X}_{1,k}$ for target $k$, the estimated location for this target is very poor, because the corresponding estimation residual is very large. In this case, for all the data association solutions $\mathcal{X}_1\in \mathcal{Y}$ that use this $\mathcal{X}_{1,k}$ to localize some target (not necessarily target $k$), we should remove them from $\mathcal{Y}$ such that they are not used for solving problem (\ref{prb:loc_with_da}) in the future. To summarize, when we implement the exhaustive search approach to solve problem (\ref{prb:da_with_loc_1}), we can keep removing some bad data association solutions in $\mathcal{Y}$, i.e., we do not need to solve problem (\ref{prb:loc_with_da}) for all the data association solutions in $\mathcal{Y}$.

\begin{figure}[t]
	\centering
	\includegraphics[height=6.6cm]{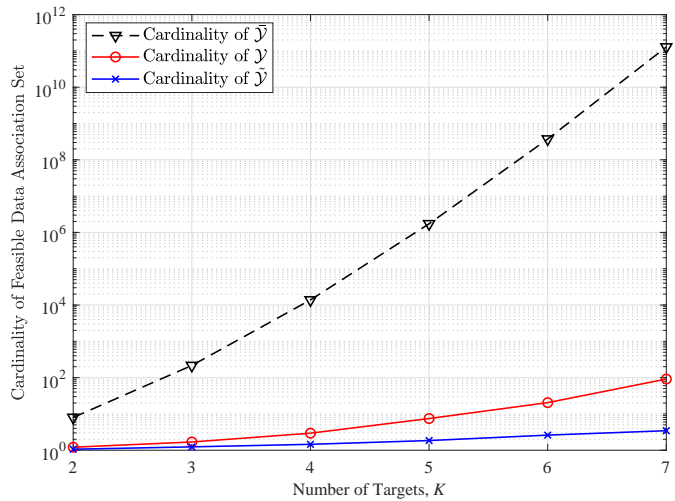}
        \vspace{-5mm}
        \caption{Comparison among the cardinalities of $\bar{\mathcal{Y}}$, $\mathcal{Y}$, and $\tilde{\mathcal{Y}}$.} \label{fig:compare_dimension_1}
        \vspace{-8mm}
\end{figure}

To verify the effectiveness of the above approach for removing the bad data association solutions in $\mathcal{Y}$, we provide a numerical example. The simulation setup is exactly the same as that for Fig. \ref{fig:compare_dimension}. For simplicity, we define $\tilde{\mathcal{Y}}$ as a subset of $\mathcal{Y}$ where all the data association solutions satisfy (\ref{eq:con_cost}) for some target $k$ are removed. Fig. \ref{fig:compare_dimension_1} shows the cardinalities of $\bar{\mathcal{Y}}$, $\mathcal{Y}$, and $\tilde{\mathcal{Y}}$ when $K$ ranges from $2$ to $7$. It is observed that via utilizing (\ref{eq:con_cost}), the number of feasible data association solutions can be further reduced. For example, when $K=7$, on average there are about $90$ data association solutions in $\mathcal{Y}$, but only about $4$ data association solutions in $\tilde{\mathcal{Y}}$. Therefore, thanks to the utilization of (\ref{eq:con_cost}), we only need to solve problem (\ref{prb:loc_with_da}) for a small number of times when solving problem (\ref{prb:da_with_loc_1}).

\begin{algorithm}[t]
	\caption{Data association and localization algorithm for single-IRS-aided networked sensing}\label{alg:da_loc_algo_single}
	    {\bf Input}: $\mathcal{D}_m^{\text{III}}$, $\mathcal{D}_m^{\text{IV}}$, $m=1,2$;\\
	    {\bf Initialization}: Set $\gamma_k=1$, $\forall k$;
     \begin{enumerate}
         \item [1.] Obtain $\mathcal{Y}$ that consists of all the feasible data association solutions satisfying \eqref{eq:con_lambda}, \eqref{eq:con_mu}, and \eqref{eq:con new}; \Comment{Step 1}
         \item [2.] Set $t=0$; {\bf Repeat}  {\bf Until} $t=|\mathcal{Y}|$: \Comment{Step 2}
         \begin{enumerate}
             \item [2.1.] Consider $\mathcal{X}_1^{(t)}\in \mathcal{Y}$, which is the $t$-th data association solution in $\mathcal{Y}$.
             \item [2.2.] Set $k=1$; {\bf Repeat} {\bf Until} $k=K$:
             \begin{enumerate}
                 \item  Given $\mathcal{X}_{1,k}^{(t)}$, which is the data association solution of target $k$ in $\mathcal{X}_1^{(t)}$, solve problem \eqref{prb:loc_with_da} via the Gauss-Newton method to obtain target $k$'s location $(x_k^{\text{T}}(\mathcal{X}_{1,k}^{(t)}),x_k^{\text{T}}(\mathcal{X}_{1,k}^{(t)}))$;
                 \item  If \eqref{eq:con_cost} is satisfied, remove all $\mathcal{X}_1$'s that use $\mathcal{X}_{1,k}^{(t)}$ as the data association solution for some target from $\mathcal{Y}$;
                 \item Set $k=k+1$;
             \end{enumerate}
             \item [2.3] Set $t=t+1$;
         \end{enumerate}
     \end{enumerate}
        {\bf{Output}}: Set the data association solution in $\mathcal{Y}$ that leads to the optimal objective value of problem (\ref{prb:loc_with_da}) as the optimal data association solution, which is denoted as $\mathcal{X}_1^{\ast}$. Given this data association solution, set the solution to problem (\ref{prb:loc_with_da}) as the optimal localization solution, which is denoted as  $\mathcal{X}_2^{\ast}$. \Comment{Step 3}
\end{algorithm}

The algorithm to solve problem (\ref{prb:da_with_loc_1}) for jointly estimating the data association variables and the localization variables under the case of one IRS is summarized in Algorithm \ref{alg:da_loc_algo_single}. It is well-known that the conventional trilateration method works in the case with at least three active anchors, under which the distances from each target to these anchors can be easily estimated. Interestingly, in this paper, we show that the trilateration method can be generalized to the scenario with two active anchors, i.e., BSs, and one passive anchor, i.e., IRS. This is because the distance between a passive target and the passive IRS can be indirectly estimated as shown by (\ref{eq:est_d_it}). In practice, the IRSs are of lower cost and can be deployed at more sites compared to the BSs. Therefore, it is practically appealing to achieve the goal of networked sensing with the aid of the IRSs. However, one limitation of single-IRS-assisted networked sensing is that this architecture can only localize the targets that are close to the IRS such that their reflected signals via the IRS is strong enough to be detected by the BSs. Therefore, in the next subsection, we aim to generalize our proposed algorithm to the multi-IRS-assisted networked sensing scenario, where multiple IRSs are deployed to enhance the coverage region such that each target at any position is close to some IRS.
\vspace{-5mm}
\subsection{Data Association and Localization with Multiple IRSs}\label{sec:da_loc_multi}
In the case with multiple IRSs, the new challenge to solve problem (\ref{prb:da_with_loc_1}) is that the mapping between each target $k$ and its associated IRS, i.e., $\gamma_k$, is unknown, $\forall k$. One straightforward approach to tackle the issue arising from $\gamma_k$'s is based on exhaustive search. Specifically, we can find all the data association solutions that satisfy (\ref{eq:con_lambda}), (\ref{eq:con_mu}), (\ref{eq:con gamma}), and (\ref{eq:con new}) to get $\mathcal{Y}$. Then, with the knowledge of $\mathcal{Y}$, we can apply Algorithm \ref{alg:da_loc_algo_single} to estimate the locations of the targets. However, the above approach is of very high complexity in the case with multiple IRSs. This is because in this case, there are a huge number of target-IRS association solutions that satisfy (\ref{eq:con gamma}). In other words, the cardinality of $\mathcal{Y}$ is significantly increased with multiple IRSs.

To tackle the above issue, we aim to reveal some hidden property about the feasible data association solutions for reducing their number in the case with multiple IRSs. Our key point is that if the distance between any two IRSs is very large, there is no overlap among the coverage regions of different IRSs, and each target should be associated with its closest IRS. In the following, we show how to utilize the above property to significantly reduce the number of feasible data association solutions.

Specifically, given any $\lambda_{1,k}$ and $\lambda_{2,k}$ for some target $k$, the corresponding estimated location for target $k$ should satisfy
\begin{align}
    &\sqrt{(x_{k}^{\text{T}}-x_1^{\text{B}})^2+(y_{k}^{\text{T}}-y_1^{\text{B}})^2}=\frac{1}{2}\mathcal{D}_1^{\text{III}}(\lambda_{1,k}),\label{eq:circle_1}\\
    &\sqrt{(x_{k}^{\text{T}}-x_2^{\text{B}})^2+(y_{k}^{\text{T}}-y_2^{\text{B}})^2}=\frac{1}{2}\mathcal{D}_2^{\text{III}}(\lambda_{2,k}).\label{eq:circle_2}
\end{align}Note that there are either no solutions or two (same or different) solutions for ($x_{k}^{\text{T}},y_{k}^{\text{T}}$) to the above equations. If there are two location solutions for target $k$ to the above equations given some $\lambda_{1,k}$ and $\lambda_{2,k}$, define $(x_{k,i}^{\text{T}}(\lambda_{1,k},\lambda_{2,k}),y_{k,i}^{\text{T}}(\lambda_{1,k},\lambda_{2,k}))$ as the $i$-th solution, $i=1,2$, and $\mathcal{J}_k(\lambda_{1,k},\lambda_{2,k})=\{(x_{k,1}^{\text{T}}(\lambda_{1,k},\lambda_{2,k}),y_{k,1}^{\text{T}}(\lambda_{1,k},\lambda_{2,k})),(x_{k,2}^{\text{T}}(\lambda_{1,k},\lambda_{2,k}),y_{k,2}^{\text{T}}(\lambda_{1,k},\lambda_{2,k}))\}$ as the set that consists of these two solutions. For the $i$-th solution to (\ref{eq:circle_1}) and (\ref{eq:circle_2}), if it is the true location of target $k$, then the associated IRS to target $k$, which should be closest to target $k$, is given as
\begin{align}\label{prb:optimal_irs}
    \gamma_{k,i}(\lambda_{1,k},\lambda_{2,k})=\text{arg}~\underset{r}{\text{min}}~\sqrt{(x_{k,i}^{\text{T}}(\lambda_{1,k},\lambda_{2,k})-x_{r}^{I})^2+(y_{k,i}^{\text{T}}(\lambda_{1,k},\lambda_{2,k})-y_{r}^{I})^2}, ~ i=1,2.
\end{align}It is observed that given any $\lambda_{1,k}$ and $\lambda_{2,k}$ for some target $k$, there are at most two possible values of $\gamma_k$, i.e., $\gamma_{k,1}(\lambda_{1,k},\lambda_{2,k})$ and $\gamma_{k,2}(\lambda_{1,k},\lambda_{2,k})$. If we do not utilize the property that each target is associated with its closest IRS, given any $\lambda_{1,k}$ and $\lambda_{2,k}$, there are $R$ possible values of $\gamma_k$ as shown in (\ref{eq:con gamma}).

\begin{figure}[t]
	\centering
	\includegraphics[height=6.6cm]{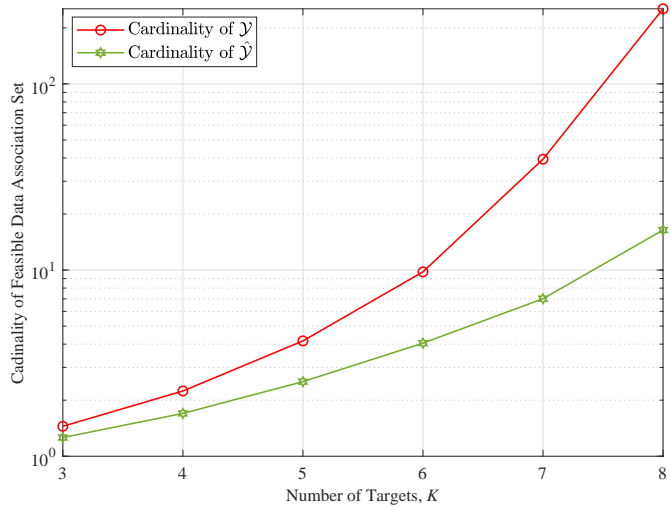}
    \caption{Comparison among the cardinalities of $\hat{\mathcal{Y}}$ and $\mathcal{Y}$ with $3$ IRSs.} \label{fig:compare_dimension_multi}
    \vspace{-8mm}
\end{figure}

To summarize, via utilizing (\ref{prb:optimal_irs}), we can define a new set for the feasible data association solutions in the case with multiple IRSs as
\begin{align}\label{eqn:new data association}
\hat{\mathcal{Y}}=\{\mathcal{X}_1|\mathcal{X}_1\in \mathcal{Y} ~ \text{and} ~ \gamma_k\in \mathcal{B}_k(\lambda_{1,k},\lambda_{2,k}), ~ \forall k,\},
\end{align}where $\mathcal{B}_k(\lambda_{1,k},\lambda_{2,k})=\{\gamma_{k,1}(\lambda_{1,k},\lambda_{2,k}),\gamma_{k,2}(\lambda_{1,k},\lambda_{2,k})\}$. In the following, we provide one numerical example to verify the effectiveness of the above approach to reduce the cardinality of the set $\mathcal{Y}$ under the multi-IRS case. In this example, we assume there are $R=3$ IRSs located at $(-60,60)$, $(70,60)$, and $(0,-70)$ in meter. Other simulation setups are the same as that of Fig. \ref{fig:compare_dimension}. Fig. \ref{fig:compare_dimension_multi} shows the cardinalities of $\mathcal{Y}$ and $\hat{\mathcal{Y}}$ for different values of $K$, ranging from $3$ to $8$. It is observed that the utilization of \eqref{prb:optimal_irs} significantly reduces the number of potential data association solutions. For example, when $K=8$, there are about $250$ data association solutions in $\mathcal{Y}$, but only $18$ data association solutions in $\hat{\mathcal{Y}}$. This thus verifies the effectiveness of the above approach.

After obtaining the set $\hat{\mathcal{Y}}$, we may follow Algorithm \ref{alg:da_loc_algo_single} to solve problem (\ref{prb:da_with_loc_1}) in the case with multiple IRSs. The algorithm is summarized in Algorithm \ref{alg:da_loc_algo_multi}.

\begin{algorithm}[t]
	\caption{Data association and localization algorithm for multi-IRS-aided networked sensing}\label{alg:da_loc_algo_multi}
	    \textbf{Input}: $\mathcal{D}_m^{\text{III}}$, $\mathcal{D}_m^{\text{IV}}$, $m=1,2$;
     \begin{enumerate}
         \item[1.] Obtain $\hat{\mathcal{Y}}$ based on (\ref{eqn:new data association});
         \item[2.] Implement Step $2$ of Algorithm \ref{alg:da_loc_algo_single}, where $\mathcal{Y}$ is replace by $\hat{\mathcal{Y}}$;
     \end{enumerate}
    {\bf{Output}}: Set the data association solution in $\hat{\mathcal{Y}}$ that leads to the optimal objective value of problem \eqref{prb:loc_with_da} as the optimal data association solution, which is denoted as $\mathcal{X}_1^{\ast}$. Given this data association solution, set the solution to problem \eqref{prb:loc_with_da} as the optimal localization solution, which is denoted as $\mathcal{X}_2^{\ast}$.
\end{algorithm}
\vspace{-3mm}
\section{Numerical Results}\label{sec:numerical}

In this section, we provide numerical results to verify the effectiveness of the proposed two-phase localization protocol for IRS-enabled networked sensing. In these numerical examples, the channel bandwidth is $400$ MHz. The identical transmit power of BS $1$ and BS $2$ is $39$ dBm. The power spectrum density of the noise at the BS is $-174$ dBm/Hz. It is assumed that BS $1$ and BS $2$ are located at $(100, 0)$ and $(-100,0)$ in meter, respectively. Moreover, given the location of each IRS, some targets are randomly located within a semi-circle whose center is this IRS and radius is $50$ meters. To evaluate the performance of the proposed scheme, we consider two benchmark schemes. Under the first benchmark scheme, we assume that the data association solutions of $\gamma_k$'s, $\lambda_{m,k}$'s, and $\mu_{m,k}$'s are perfectly known. This scheme can provide a performance upper bound to evaluate the effectiveness of the proposed algorithms. Under the second benchmark scheme, we replace the passive IRSs with the active micro BSs that can transmit/receive radio signals to/from the targets within their coverage region (note that BS $1$ and BS $2$ can cover all the targets). In this scheme, we can apply the algorithm proposed in \cite{dvc} for localization.

\begin{figure}[t]
	\centering
	\subfigure[Localization performance]{\includegraphics[height=6cm]{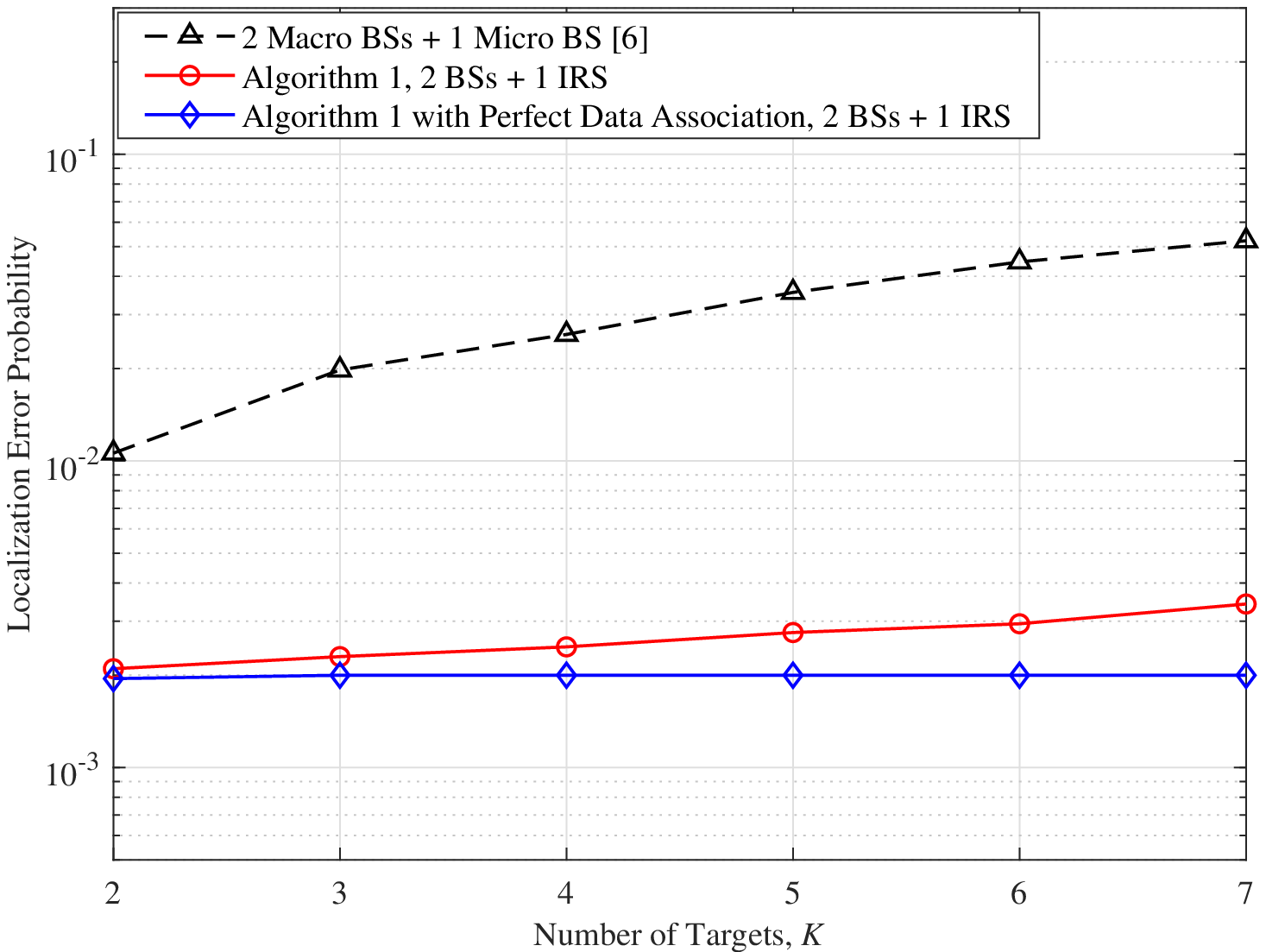}\label{fig:perf_single}}
	\subfigure[CPU Time]{\includegraphics[height=6cm]{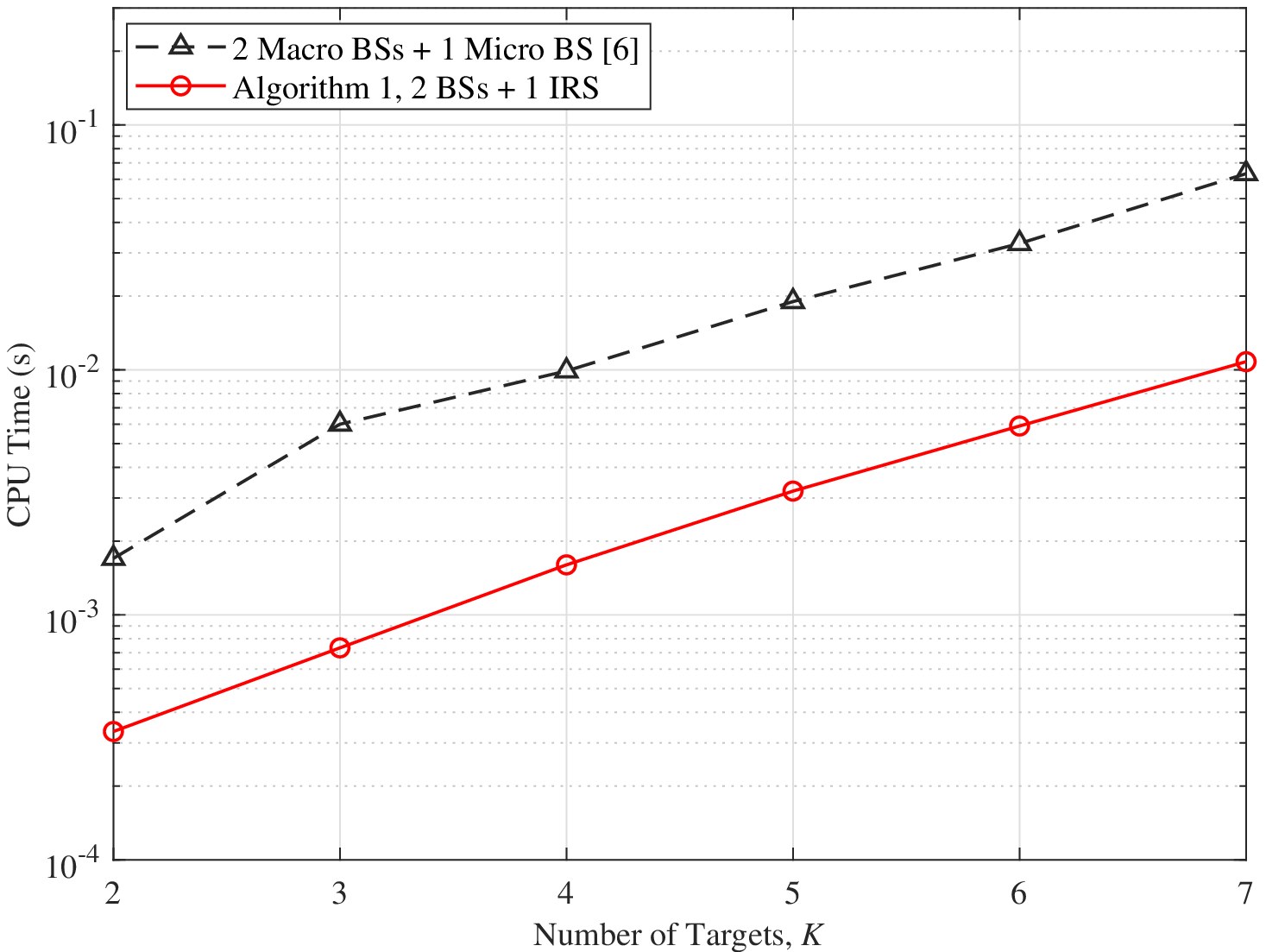}\label{fig:time_single}}
    \caption{Evaluation of Algorithm \ref{alg:da_loc_algo_single} under a single-IRS-aided device-free sensing system.} \label{fig:all_single}
    \vspace{-8mm}
\end{figure}
\vspace{-5mm}
\subsection{Data Association and Localization Performance with One IRS}
First, we show the numerical results under the single-IRS case where the IRS is located at $(0,40)$ in meter. Fig. \ref{fig:perf_single} and Fig. \ref{fig:time_single} show the performance comparison between our proposed Algorithm \ref{alg:da_loc_algo_single} and the two benchmark schemes, in terms of localization accuracy and computational complexity, respectively, when the number of targets ranges from $2$ to $7$. First, Fig. \ref{fig:perf_single} shows the localization error probabilities achieved by Algorithm \ref{alg:da_loc_algo_single} and two benchmark schemes. Here, an error event for localizing a target is defined as the case that the estimated location is not lying within a radius of $0.8$ meter from the true target location. In this numerical example, we generate $10^5$ realizations of the targets' locations and record the total number of error events as $N_{\text{error}}$. Then, the localization error probability is defined as $\frac{N_\text{error}}{K\times 10^5}$. It is observed that under Algorithm \ref{alg:da_loc_algo_single}, the localization error probability is very low. Moreover, it is observed that the localization error probability under IRS-enabled networked sensing is lower than that under the conventional networked sensing strategy with $3$ BSs. This is because the utilization of \eqref{eq:con new} significantly reduces the cardinality of the set that contains all the feasible data association solutions, as shown in Fig. \ref{fig:compare_dimension}, and it is more likely to find the optimal data association solution from a set with a smaller size. Last, it is observed that the localization error probability under Algorithm \ref{alg:da_loc_algo_single} is already very close to the performance upper bound, which is achieved with perfect knowledge of the data association solution.

Second, Fig. \ref{fig:time_single} shows the average CPU running time in terms of seconds to implement Algorithm \ref{alg:da_loc_algo_single} and the algorithm proposed in \cite{dvc} that utilizes $3$ BSs as anchors. It is observed that Algorithm \ref{alg:da_loc_algo_single} is of much lower computational complexity thanks to the utilization of \eqref{eq:con new} to reduce the size of the set that consists of the feasible data association solutions.

\begin{figure}[t]
	\centering
	\subfigure[Localization performance]{\includegraphics[height=6cm]{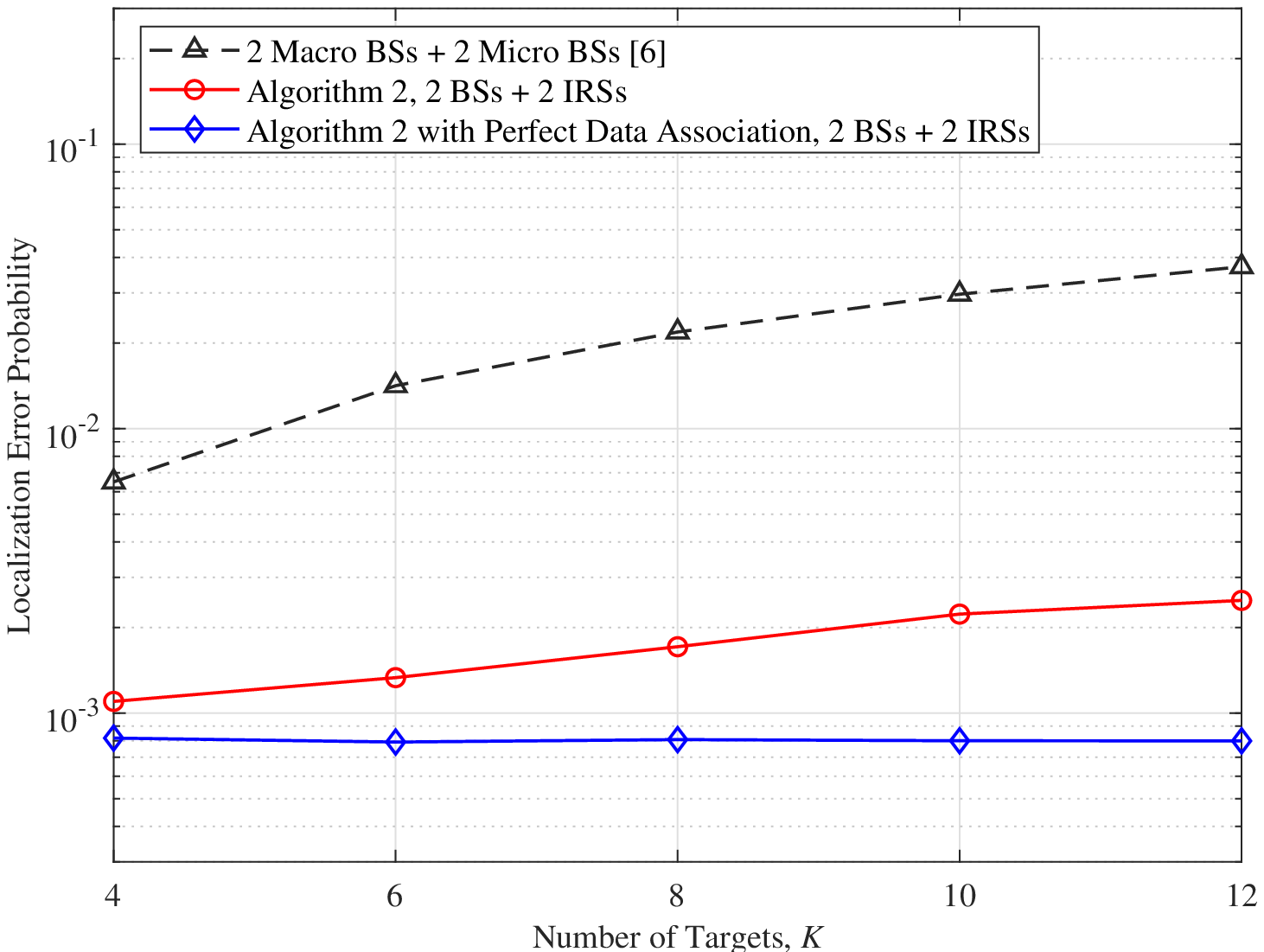}\label{fig:perf_multi}}
	\subfigure[CPU Time]{\includegraphics[height=6cm]{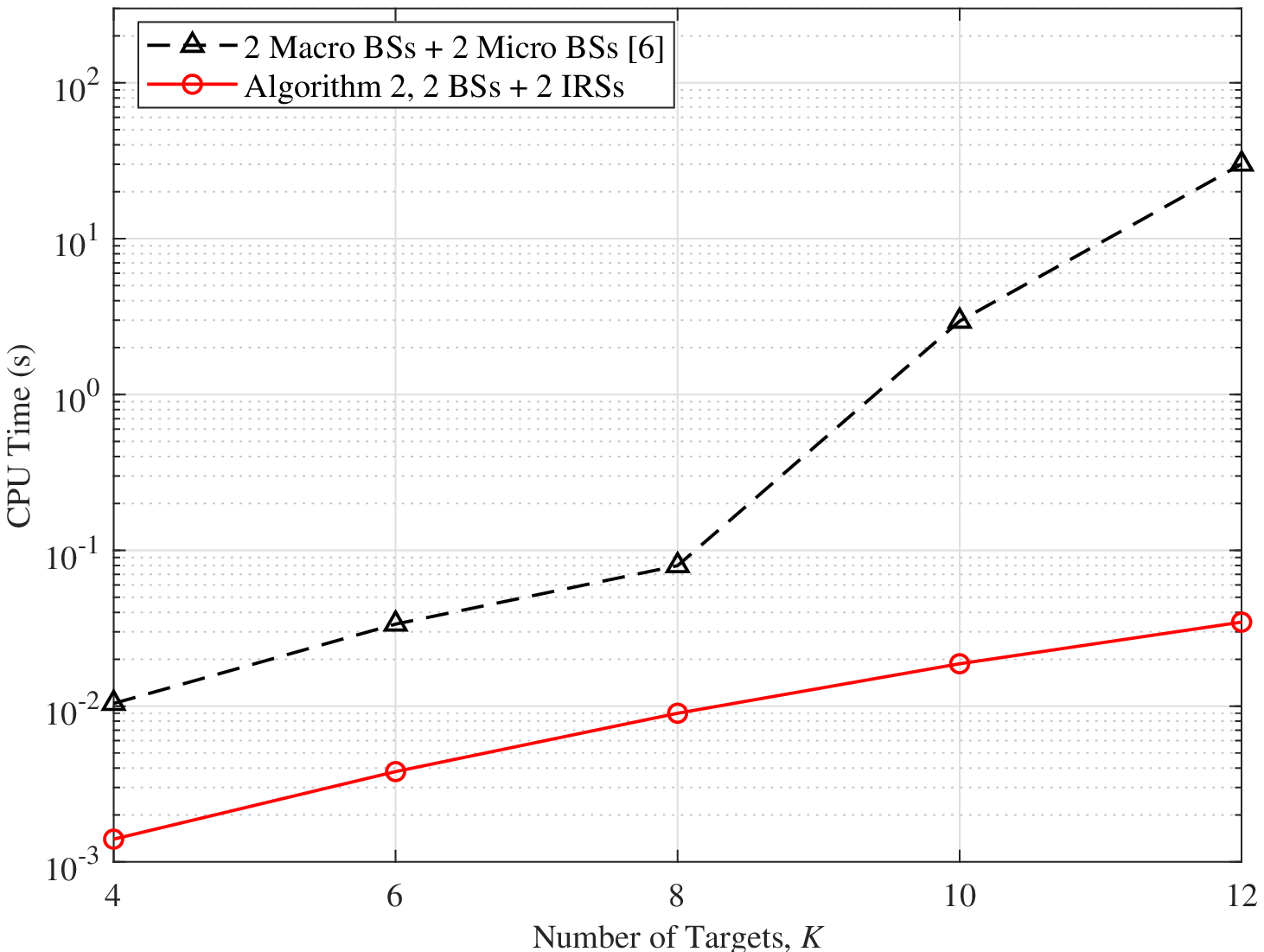}\label{fig:time_multi}}
    \caption{Evaluation of Algorithm \ref{alg:da_loc_algo_multi} under a multi-IRS-aided device-free sensing system.} \label{fig:all_multi}
    \vspace{-8mm}
\end{figure}
\vspace{-5mm}
\subsection{Data Association and Localization Performance with Multiple IRSs}
Next, we show the numerical results under the multi-IRS case. We assume that there are $2$ IRSs located at $(-60,40)$ and $(70,40)$ in meter. Fig. \ref{fig:perf_multi} and Fig. \ref{fig:time_multi} show the localization accuracy and the computational complexity of Algorithm \ref{alg:da_loc_algo_multi}, respectively, when the number of targets ranges from $4$ to $12$. It is observed from Fig. \ref{fig:perf_multi} that similar to the case with one IRS, the performance of Algorithm \ref{alg:da_loc_algo_multi} under the heterogeneous architecture is much better than that of the conventional networked sensing strategy with only BSs, and very close to the performance upper bound with perfect data association solution. It is also observed from Fig. \ref{fig:time_multi} that Algorithm \ref{alg:da_loc_algo_multi} is of much lower complexity compared to the case when all the anchors are BSs. This thus validates the feasibility and effectiveness of the proposed algorithm for multi-IRS-enabled networked sensing.

\vspace{-3mm}
\subsection{Effect of Network Topology on Localization Performance}

In practice, the locations of the anchors will significantly affect the localization accuracy. In this subsection, we evaluate the effect of the IRS deployment strategy on localization error probability. Specifically, Theorem \ref{theorem1} indicates that in the ideal case with perfect range estimation in Phase I of our proposed protocol, the correct data association solution and localization solution can be found in Phase II almost surely, as long as the locations of the IRSs satisfy conditions C$1$ and C$2$. In the following, we provide numerical results to show that if conditions C$1$ and C$2$ in Theorem \ref{theorem1} are not satisfied, the localization performance can be degraded.

\begin{figure}[t]
	\centering
	\subfigure[Effect of Condition C$1$ on Localization Accuracy]{\includegraphics[height=6cm]{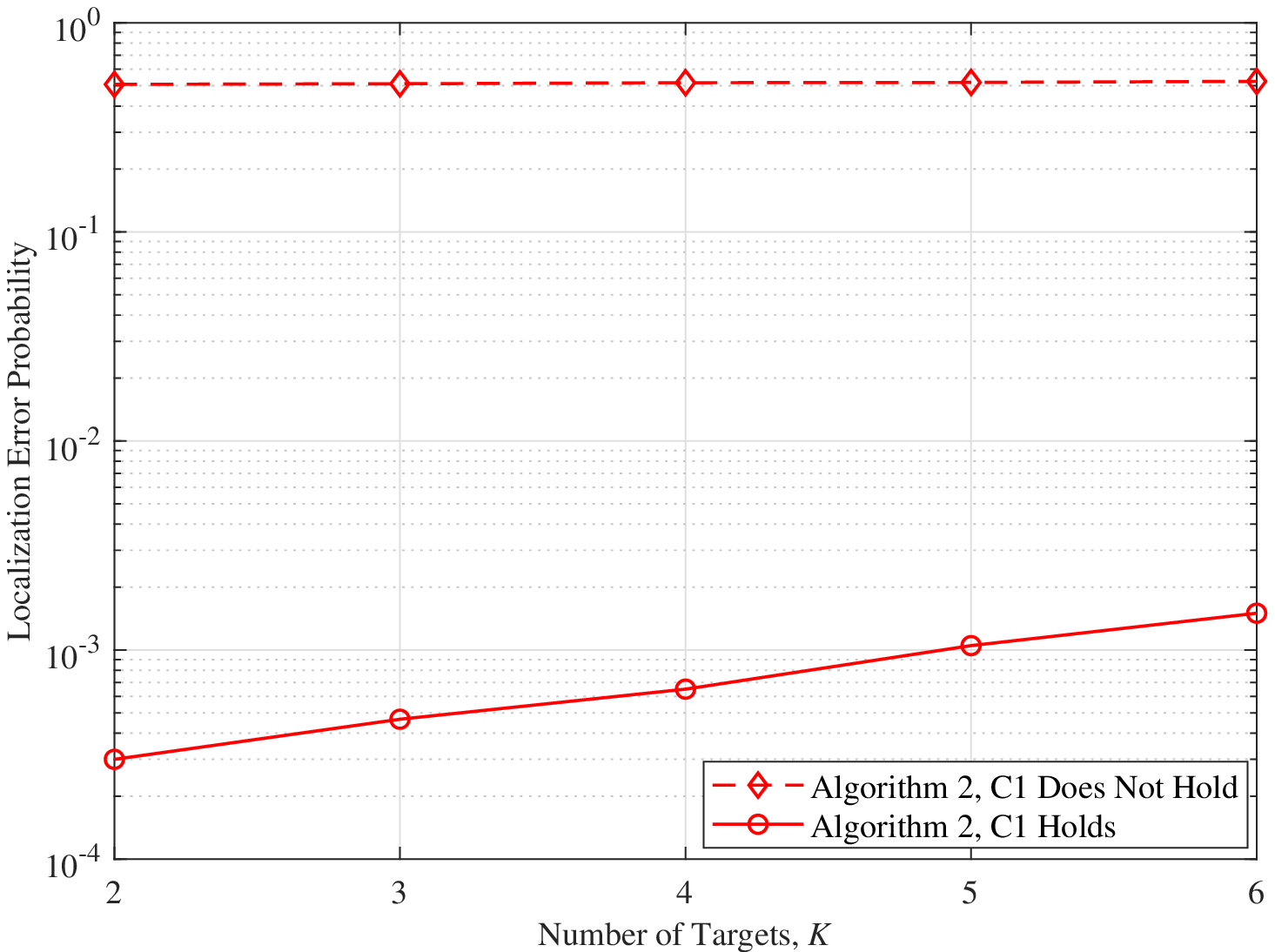}\label{fig:perf_con_c1}}
	\subfigure[Effect of Condition C$2$ on Localization Accuracy]{\includegraphics[height=6cm]{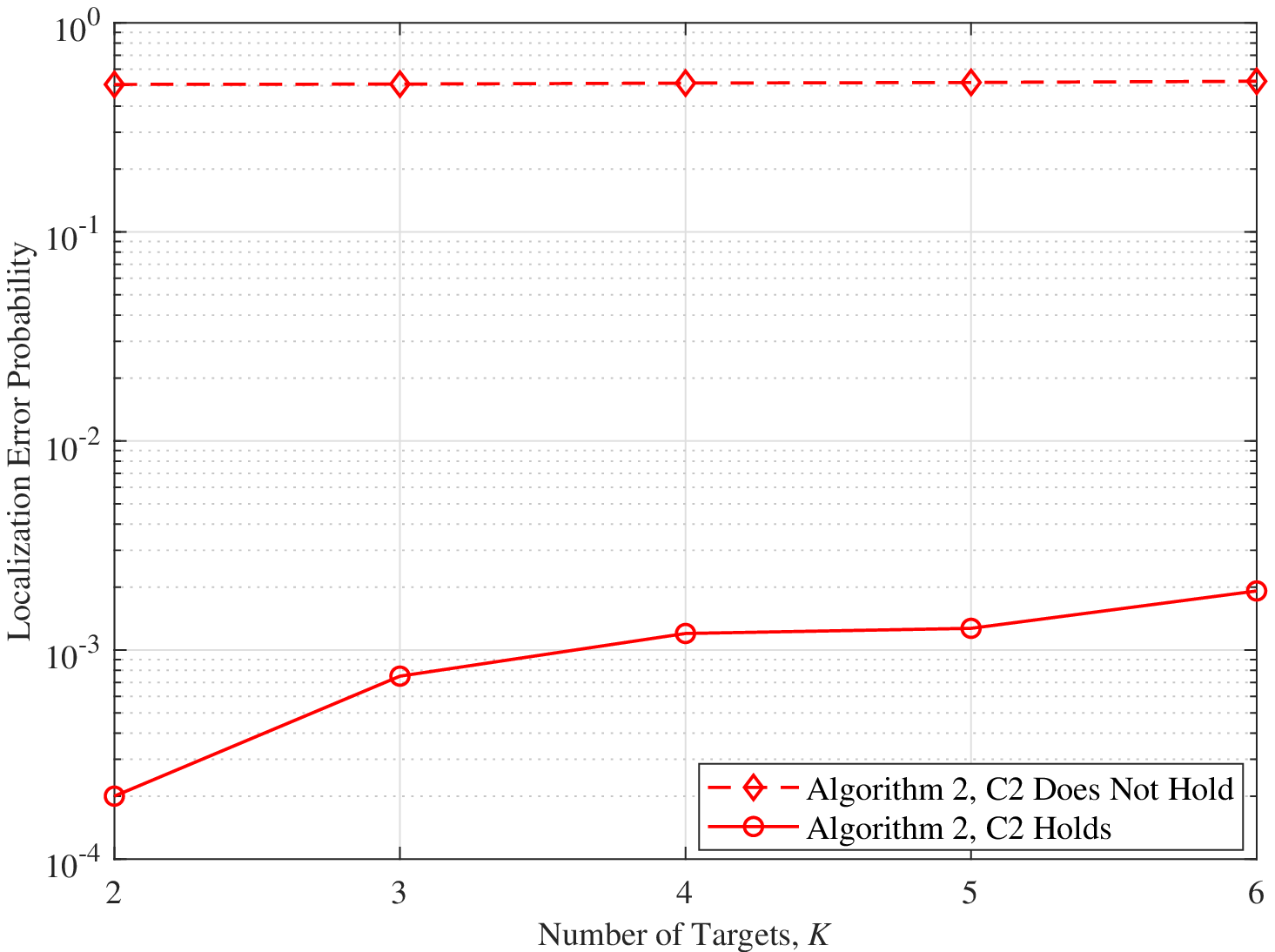}\label{fig:perf_con_c2}}
    \caption{Evaluation of Algorithm \ref{alg:da_loc_algo_multi} under different network topologies.} \label{fig:effect_topologies}
    \vspace{-8mm}
\end{figure}

To evaluate the effect of condition C$1$, we fix the first IRS's location as $(0,60)$ in meter, while the second IRS's location is set as $(0,-60)$ in meter (C$1$ does not hold) and $(30,-60)$ in meter (C$1$ holds), respectively. Fig. \ref{fig:perf_con_c1} shows the localization accuracy achieved by Algorithm \ref{alg:da_loc_algo_multi} under the above IRS deployment strategies. To evaluate the effect of condition C$2$, we fix the first IRS's location as $(80,-60)$ in meter, while the second IRS's location is set as $(80,-60)$ in meter (C$2$ does not hold) and $(120,-60)$ in meter (C$2$ holds), respectively. Fig. \ref{fig:perf_con_c2} shows the localization accuracy achieved by Algorithm \ref{alg:da_loc_algo_multi} under the above IRS deployment strategies. It is observed that the localization accuracy is not good if condition C$1$ or condition C$2$ in Theorem \ref{theorem1} does not hold. Therefore, we should carefully select the sites to deploy the IRSs to achieve good localization accuracy under our proposed heterogeneous networked sensing architecture.

\section{Conclusion}\label{sec:conclusion}
In this paper, we proposed a novel heterogeneous networked sensing architecture consisting of both the active anchors, i.e., BSs, and the passive anchors, i.e., IRSs, to enhance the anchor density with low cost. There are two main challenges for employing passive anchors under our proposed architecture. First, it is difficult to directly measure the distance between a target and its associated IRS, because both of them are passive and cannot estimate the propagation delay of the round-trip signals between them. Second, before localizing each target, we do not know which IRS is closet to it and serves as its anchor for localizing it. We proposed efficient signal processing methods to tackle the above two challenges. Numerical results showed that our proposed heterogenous networked sensing architecture can achieve the same or even better performance compared to the conventional architecture with active anchors.

\begin{appendices}
\section{Proof of Theorem \ref{theorem1}}\label{appendix1}
We prove Theorem \ref{theorem1} by showing that the data association solution satisfying \eqref{eq:con_lambda}, \eqref{eq:con_mu}, \eqref{eq:con gamma}, and \eqref{eq:con new} is unique. If this is true, then the localization solution must be unique and correct, because three anchors (that are not on one line) can uniquely localize a target. For convenience, we first define the $X$-axis of the considered 2-D Cartesian coordinate system as the line passing through BS $1$ and BS $2$. As a result, the coordinates of the two BSs are $(x_1^{\text{B}},0)$ and $(x_2^{\text{B}},0)$. We further define the $k$-th target as the target whose distance to BS $1$ is the $k$-th largest element in $\mathcal{D}_1^{\text{III}}$, i.e., $\lambda_{1,k}=k$, $\forall k$. In the rest of the proof, we aim to show that given any false data association solution $\mathcal{X}_{1,k}^\text{F}$ for localizing target $k$ that is different from the correct data association solution $\mathcal{X}_{1,k}^\text{C}$, i.e., $[k,\lambda_{2,k}^{\text{F}},\mu_{1,k}^{\text{F}},\mu_{2,k}^{\text{F}},\gamma_k^{\text{F}}]\neq [k,\lambda_{2,k}^{\text{C}},\mu_{1,k}^{\text{C}},\mu_{2,k}^{\text{C}},\gamma_k^{\text{C}}]$, the condition \eqref{eq:con new} with $\tau = 0$ does not hold almost surely, i.e.,
\begin{align}\label{eq:proof_con}
    \mathcal{D}_1^{\text{IV}}(\mu_{1,k}^{\text{F}})-\sqrt{(x_k^{\text{T}}-x_1^{\text{B}})^2+(y_k^{\text{T}}-y_1^{\text{B}})^2}-d_{1,\gamma_k^{\text{F}}}^{\text{BI}}=\mathcal{D}_2^{\text{IV}}(\mu_{2,k}^{\text{F}})-\frac{1}{2}\mathcal{D}_2^{\text{III}}(\lambda_{2,k}^{\text{F}})-d_{2,\gamma_k^{\text{F}}}^{\text{BI}},
\end{align}does not hold almost surely, $\forall k$. If this is true, then the false data association solution $\mathcal{X}_1^{\text{F}}$ for localizing all the $K$ targets does not exist almost surely, because the number of targets in the considered system is finite.

Define $\mathcal{E}$ as the event that condition \eqref{eq:proof_con} holds for false data association solution $\mathcal{X}_{1,k}^\text{F}$. We further let $\mathcal{B}$ denote the set of $(x_{1}^{\text{T}},y_{1}^{\text{T}},\cdots,x_K^{\text{T}},y_K^{\text{T}})$ such that if $(x_{1}^{\text{T}},y_{1}^{\text{T}},\cdots,x_K^{\text{T}},y_K^{\text{T}})\in\mathcal{B}$, $(x_{1}^{\text{T}},y_{1}^{\text{T}},\cdots,x_K^{\text{T}},y_K^{\text{T}})$ must satisfy \eqref{eq:proof_con}. Moreover, define $\tilde{\mathcal{B}}$ as the set of $(x_{1}^{\text{T}},y_{1}^{\text{T}},\cdots,x_{k-1}^{\text{T}},y_{k-1}^{\text{T}},$\\$x_{k+1}^{\text{T}},y_{k+1}^{\text{T}},\cdots,x_{K}^{\text{T}},y_{K}^{\text{T}})$ such that if $(x_{1}^{\text{T}},y_{1}^{\text{T}},\cdots,x_{k-1}^{\text{T}},y_{k-1}^{\text{T}},x_{k+1}^{\text{T}},y_{k+1}^{\text{T}},\cdots,x_{K}^{\text{T}},y_{K}^{\text{T}})\in\tilde{\mathcal{B}}$, there exists $(x_k^{\text{T}},y_k^{\text{T}})$ to satisfy $(x_{1}^{\text{T}},y_{1}^{\text{T}},\cdots,x_{K}^{\text{T}},y_{K}^{\text{T}})\in\mathcal{B}$, and $\hat{\mathcal{B}}$ as the set of $(x_k^{\text{T}},y_k^{\text{T}})$ such that given any $(x_{1}^{\text{T}},y_{1}^{\text{T}},\cdots,x_{k-1}^{\text{T}},y_{k-1}^{\text{T}},x_{k+1}^{\text{T}},y_{k+1}^{\text{T}},\cdots,x_{K}^{\text{T}},y_{K}^{\text{T}})\in\tilde{\mathcal{B}}$, if $(x_k^{\text{T}},y_k^{\text{T}})\in\hat{\mathcal{B}}$, then the coordinates of $(x_{1}^{\text{T}},y_{1}^{\text{T}},\cdots,x_{K}^{\text{T}},y_{K}^{\text{T}})$ satisfy \eqref{eq:proof_con}. The probability of $\mathcal{E}$ is thus expressed as
\begin{align}\label{eq:proof_pq}
&\text{Pr}(\mathcal{E})=\int_{(x_{1}^{\text{T}},y_{1}^{\text{T}},\cdots,x_K^{\text{T}},y_K^{\text{T}}) \in \mathcal{B}} p\left(x_{1}^{\text{T}},y_{1}^{\text{T}},\cdots,x_K^{\text{T}},y_K^{\text{T}}\right) d x_{1}^{\text{T}} d y_{1}^{\text{T}} \cdots d x_{K}^{\text{T}} d y_{K}^{\text{T}}\notag\\
&\overset{\text{a}}{=}\int_{(x_{1}^{\text{T}},y_{1}^{\text{T}},\cdots,x_{k-1}^{\text{T}},y_{k-1}^{\text{T}},x_{k+1}^{\text{T}},y_{k+1}^{\text{T}},\cdots,x_{K}^{\text{T}},y_{K}^{\text{T}})\in \tilde{\mathcal{B}}}\left(\int_{(x_k^{\text{T}},y_k^{\text{T}}) \in \hat{\mathcal{B}}} p\left(x_{k}^{\text{T}}, y_{k}^{\text{T}}\right) d x_{k}^{\text{T}} d y_{k}^{\text{T}}\right)\notag \\
&\times p\left(x_{1}^{\text{T}},y_{1}^{\text{T}},\cdots,x_{k-1}^{\text{T}},y_{k-1}^{\text{T}},x_{k+1}^{\text{T}},y_{k+1}^{\text{T}},\cdots,x_{K}^{\text{T}},y_{K}^{\text{T}}\right) d x_{1}^{\text{T}} d y_{1}^{\text{T}}\cdots d x_{k-1}^{\text{T}} d y_{k-1}^{\text{T}}d x_{k+1}^{\text{T}} d y_{k+1}^{\text{T}}\cdots d x_{K}^{\text{T}} d y_{K}^{\text{T}},
\end{align}where $p(\cdot)$ is the probability density function (PDF) and step (a) is because all the targets are independently located in the considered system. In the following, we show that $\hat{\mathcal{B}}$ is at most a line given any false data association solution $\mathcal{X}_{1,k}^\text{F} \neq \mathcal{X}_{1,k}^\text{C}$. If this is true, then $\text{Pr}(\mathcal{E})=0$ holds almost surely, $\forall \mathcal{X}_{1,k}^\text{F} \neq \mathcal{X}_{1,k}^\text{C}$, because target $k$ is uniformly distributed in the network such that
\begin{align}\label{eq:proof_bf_as0}
   \int_{\left(x_{k}^{\text{T}}, y_{k}^{\text{T}}\right) \in \hat{\mathcal{B}}} p\left(x_{k}^{\text{T}}, y_{k}^{\text{T}}\right) d x_{k}^{\text{T}} d y_{k}^{\text{T}}\overset{a.s.}{=}0.
\end{align}

In the following, we discuss the set $\hat{\mathcal{B}}$ given different $\mathcal{X}_{1,k}^{\text{F}}\neq \mathcal{X}_{1,k}^{\text{C}}$.\\
\textbf{Case I}: the data association variable $\gamma_k^{\text{F}}$ is false, and the data association variables $[\lambda_{2,k}^{\text{F}},\mu_{1,k}^{\text{F}},\mu_{2,k}^{\text{F}}]$ are correct. Under \textbf{Case I}, because the data association variables $[\lambda_{2,k}^{\text{F}},\mu_{1,k}^{\text{F}},\mu_{2,k}^{\text{F}}]$ are correct, i.e., $[\lambda_{2,k}^{\text{F}},\mu_{1,k}^{\text{F}},\mu_{2,k}^{\text{F}}]=[\lambda_{2,k}^{\text{C}},\mu_{1,k}^{\text{C}},\mu_{2,k}^{\text{C}}]$,
\eqref{eq:proof_con} reduces to
\begin{align}
    \sqrt{\!(x_{\gamma_k^{\text{C}}}^{\text{I}}\!-\!x_{2}^{\text{B}})^2\!+\!(y_{\gamma_k^{\text{C}}}^{\text{I}})^2}\!-\!\sqrt{\!(x_{\gamma_k^{\text{F}}}^{\text{I}}\!-\!x_2^{\text{B}})^2\!+\!(y_{\gamma_k^{\text{F}}}^{\text{I}})^2}\!
=\!\sqrt{\!(x_{\gamma_k^{\text{C}}}^{\text{I}}\!-\!x_{1}^{\text{B}})^2\!+\!(y_{\gamma_k^{\text{C}}}^{\text{I}})^2}\!-\!\sqrt{(x_{\gamma_k^{\text{F}}}^{\text{I}}\!-\!x_1^{\text{B}})^2\!+\!(y_{\gamma_k^{\text{F}}}^{\text{I}})^2},
\end{align}which contradicts with the assumption that any two IRSs $\gamma_k^{\text{C}}$ and $\gamma_k^{\text{F}}$ satisfy $d_{2,\gamma_k^{\text{C}}}^{\text{BI}}-d_{1,\gamma_k^{\text{C}}}^{\text{BI}}\neq d_{2,\gamma_k^{\text{F}}}^{\text{BI}}-d_{1,\gamma_k^{\text{F}}}^{\text{BI}}$, $\forall \gamma_k^{\text{C}}\ne \gamma_k^{\text{F}}$. Thus, $\hat{\mathcal{B}}$ is an empty set under \textbf{Case I}.\\
\textbf{Case II}: the data association variable $\gamma_k^{\text{F}}$ is false, and two out of the three data association variables $[\lambda_{2,k}^{\text{F}},\mu_{1,k}^{\text{F}},\mu_{2,k}^{\text{F}}]$ are correct.\footnote{Note that if two out of three variables $[\lambda_{2,k}^{\text{F}},\mu_{1,k}^{\text{F}},\mu_{2,k}^{\text{F}}]$ are correct, the variable $\gamma_k^{\text{F}}$ must be false. For example, suppose $\lambda_{2,k}^{\text{F}}$ and $\mu_{1,k}^{\text{F}}$ are correct, then according to \eqref{eq:con new}, $\mu_{2,k}^{\text{F}}$ must be correct given a correct $\gamma_k^{\text{F}}$. This contradicts with the assumption that two out of $[\lambda_{2,k}^{\text{F}},\mu_{1,k}^{\text{F}},\mu_{2,k}^{\text{F}}]$ are correct.} Without loss of generality, we focus on the event that $\lambda_{2,k}^{\text{F}}$ and $\mu_{1,k}^{\text{F}}$ are correct while $\mu_{2,k}^{\text{F}}$ is false. Other sub-events, e.g., $\mu_{1,k}^{\text{F}}$ and $\mu_{2,k}^{\text{F}}$ are correct while $\lambda_{2,k}^{\text{F}}$ is false, can be shown using a similar approach. Under this case, \eqref{eq:proof_con} reduces to
\begin{align}\label{eq:proof_caseii}
    \sqrt{\!(x_k^{\text{T}}\!-\!x_1^{\text{B}})^2\!+\!(y_k^{\text{T}})^2}\!+\!\sqrt{\!(x_k^{\text{T}}\!-\!x_2^{\text{B}})^2\!+\!(y_k^{\text{T}})^2}\!+\!\sqrt{\!(x_k^{\text{T}}\!-\!x_{\gamma_k^{\text{C}}}^{\text{I}})^2\!+\!(y_k^{\text{T}}\!-\!y_{\gamma_k^{\text{C}}}^{\text{I}})^2} \!=\! \mathcal{D}_2^{\text{IV}}(\mu_{2,k}^\text{F})\!+\!d_{1,\gamma_{k}^{\text{F}}}^{\text{BI}}\!-\!d_{2,\gamma_{k}^{\text{F}}}^{\text{BI}}.
\end{align}Note that given $(x_{1}^{\text{T}},y_{1}^{\text{T}},\cdots,x_{k-1}^{\text{T}},y_{k-1}^{\text{T}},x_{k+1}^{\text{T}},y_{k+1}^{\text{T}},\cdots,x_{K}^{\text{T}},y_{K}^{\text{T}})\in \tilde{\mathcal{B}}$, the right-hand side of \eqref{eq:proof_caseii} is a fixed number. If the fixed number is no larger than $0$, then $\hat{\mathcal{B}}$ is an empty set. Otherwise, $\hat{\mathcal{B}}$ is at most a set consisting of points on a 3-ellipse with fixed three foci \cite{sekino}, i.e., $(x_1^{\text{B}},0)$, $(x_2^{\text{B}},0)$, and $(x_{\gamma_{k}^{\text{C}}}^{\text{I}},y_{\gamma_{k}^{\text{C}}}^{\text{I}})$, and fixed distance sum. Thus, $\hat{\mathcal{B}}$ is at most a set consisting of points on a 3-ellipse.\\
\textbf{Case III}: the data association variable $\gamma_k^{\text{F}}$ may be correct or false, and one out of the three data association variables $[\lambda_{2,k}^{\text{F}},\mu_{1,k}^{\text{F}},\mu_{2,k}^{\text{F}}]$ is correct. Similar to \textbf{Case II}, we show one example when the variable $\lambda_{2,k}^{\text{F}}$ is correct while the others are wrong. Others can be shown to be the same similarly. When the variable $\lambda_{2,k}^{\text{F}}$ is correct while the others are wrong, \eqref{eq:proof_con} reduces to
\begin{align}\label{eq:proof_caseiii}
    \sqrt{(x_{k}^{\text{T}}-x_{2}^{\text{B}})^2+(y_{k}^{\text{T}})^2}-\sqrt{(x_{k}^{\text{T}}-x_{1}^{\text{B}})^2+(y_{k}^{\text{T}})^2}={\mathcal{D}}_2^{\text{IV}}({\mu}_{2,k}^{\text{F}})-d_{2,\gamma_k^{\text{F}}}^{\text{BI}}-\frac{1}{2}{\mathcal{D}}_1^{\text{III}}({\mu}_{1,k}^{\text{F}})+d_{1,\gamma_k^{\text{F}}}^{\text{BI}}.
\end{align}According to \eqref{eq:proof_caseiii}, given $(x_{1}^{\text{T}},y_{1}^{\text{T}},\cdots,x_{k-1}^{\text{T}},y_{k-1}^{\text{T}},x_{k+1}^{\text{T}},y_{k+1}^{\text{T}},\cdots,x_{K}^{\text{T}},y_{K}^{\text{T}})\in \tilde{\mathcal{B}}$, $\hat{\mathcal{B}}$ is either a set consisting of points on the perpendicular bisector of the line segment connecting two BSs or a set consisting of points on a given hyperbola with two BSs as foci.\\
\textbf{Case IV}: the data association variable $\gamma_k^{\text{F}}$ may be correct or false, and all of the three data association variables $[\lambda_{2,k}^{\text{F}},\mu_{1,k}^{\text{F}},\mu_{2,k}^{\text{F}}]$ are false. Under \textbf{Case IV}, \eqref{eq:proof_con} reduces to
\begin{align}\label{eq:proof_caseiv}
    \sqrt{(x_{k}^{\text{T}}-x_{1}^{\text{B}})^2+(y_{k}^{\text{T}})^2}= {\mathcal{D}}_1^{\text{IV}}({\mu}_{1,k}^{\text{F}})-d_{1,\gamma_k^{\text{F}}}^{\text{BI}}-{\mathcal{D}}_2^{\text{IV}}({\mu}_{2,k}^{\text{F}})+{\frac{1}{2}}\mathcal{D}_2^{\text{III}}({\lambda}_{2,k}^{\text{F}})+d_{2,\gamma_k^{\text{F}}}^{\text{BI}}.
\end{align}According to \eqref{eq:proof_caseiv}, given $(x_{1}^{\text{T}},y_{1}^{\text{T}},\cdots,x_{k-1}^{\text{T}},y_{k-1}^{\text{T}},x_{k+1}^{\text{T}},y_{k+1}^{\text{T}},\cdots,x_{K}^{\text{T}},y_{K}^{\text{T}})\in \tilde{\mathcal{B}}$, $\hat{\mathcal{B}}$ is either an empty set or a set consisting of points on a circle.

To summarize, given any $\mathcal{X}_{1,k}^{\text{F}} \neq \mathcal{X}_{1,k}^{\text{C}}$, $\hat{\mathcal{B}}$ is at most a line. As a result, according to \eqref{eq:proof_bf_as0}, $\text{Pr}(\mathcal{E})=0$ holds almost surely, $\forall \mathcal{X}_{1,k}^{\text{F}} \neq \mathcal{X}_{1,k}^{\text{C}}$. Because the number of IRSs and targets is finite such that the number of $\mathcal{X}_{1,k}^{\text{F}}$ is finite in the network, the false data association solution $\mathcal{X}_1^{\text{F}}$ for localizing all the $K$ targets does not exist almost surely. Theorem \ref{theorem1} is thus proved.
\end{appendices}

\vspace{-5pt}
\bibliographystyle{IEEEtran}
\bibliography{ref}

\begin{thebibliography}{10}
\providecommand{\url}[1]{#1}
\csname url@samestyle\endcsname
\providecommand{\newblock}{\relax}
\providecommand{\bibinfo}[2]{#2}
\providecommand{\BIBentrySTDinterwordspacing}{\spaceskip=0pt\relax}
\providecommand{\BIBentryALTinterwordstretchfactor}{4}
\providecommand{\BIBentryALTinterwordspacing}{\spaceskip=\fontdimen2\font plus
\BIBentryALTinterwordstretchfactor\fontdimen3\font minus
  \fontdimen4\font\relax}
\providecommand{\BIBforeignlanguage}[2]{{%
\expandafter\ifx\csname l@#1\endcsname\relax
\typeout{** WARNING: IEEEtran.bst: No hyphenation pattern has been}%
\typeout{** loaded for the language `#1'. Using the pattern for}%
\typeout{** the default language instead.}%
\else
\language=\csname l@#1\endcsname
\fi
#2}}
\providecommand{\BIBdecl}{\relax}
\BIBdecl

\bibitem{globecom22}
Q.~Wang, L.~Liu, S.~Zhang, and F.~C. Lau, ``{Trilateration-based device-free
  sensing: Two base stations and one passive IRS are sufficient},'' in
  \emph{Proc. IEEE Global Commun. Conf. (Globecom)}, Dec. 2022, pp. 5613--5618.

\bibitem{isac_survey1}
F.~Liu, C.~Masouros, A.~P. Petropulu, H.~Griffiths, and L.~Hanzo, ``Joint radar
  and communication design: Applications, state-of-the-art, and the road
  ahead,'' \emph{IEEE Trans. Commun.}, vol.~68, no.~6, pp. 3834--3862, Jun.
  2020.

\bibitem{isac_survey2}
L.~Zheng, M.~Lops, Y.~C. Eldar, and X.~Wang, ``{Radar and communication
  coexistence: An overview: A review of recent methods},'' \emph{IEEE Signal
  Process. Mag.}, vol.~36, no.~5, pp. 85--99, Sep. 2019.

\bibitem{isac_survey3}
A.~Liu \emph{et~al.}, ``{A survey on fundamental limits of integrated sensing
  and communication},'' \emph{IEEE Commun. Surveys Tuts}, vol.~24, no.~2, pp.
  994--1034, 2nd Quat. 2019.

\bibitem{Tan21}
D.~K.~P. Tan, J.~He, Y.~Li, A.~Bayesteh, Y.~Chen, P.~Zhu, and W.~Tong,
  ``{Integrated sensing and communication in 6G: Motivations, use cases,
  requirements, challenges and future directions},'' in \emph{Proc. 2021 1st
  IEEE Int. Online Symp. on Joint Commun. $\&$ Sens. (JC$\&$S)}, Feb. 2021.

\bibitem{dvc}
Q.~Shi, L.~Liu, S.~Zhang, and S.~Cui, ``{Device-free sensing in OFDM cellular
  network},'' \emph{IEEE J. Sel. Areas Commun.}, vol.~40, no.~6, pp.
  1838--1853, Jun. 2022.

\bibitem{Mahler07}
R.~Mahler, \emph{Statistical Multisource-Multitarget Information Fusion},
  Norwood, MA, USA: Artech House, 2007.

\bibitem{irs_survey1}
Q.~Wu, S.~Zhang, B.~Zheng, C.~You, and R.~Zhang, ``{Intelligent reflecting
  surface-aided wireless communications: A tutorial},'' \emph{IEEE Trans.
  Commun.}, vol.~69, no.~5, pp. 3313--3351, May.

\bibitem{irs_survey2}
M.~D. Renzo \emph{et~al.}, ``{Smart radio environments empowered by
  reconfigurable AI meta-surfaces: An idea whose time has come},''
  \emph{EURASIP J. Wireless Commun. Network}, no. 129, pp. 1--20, May 2019.

\bibitem{irs_loc_mag}
J.~He, F.~Jiang, K.~Keykhosravi, J.~Kokkoniemi, H.~Wymeersch, and M.~Juntti,
  ``{Beyond 5G RIS mmWave systems: Where communication and localization
  meet},'' \emph{IEEE Access}, vol.~10, pp. 68\,075--68\,084, 2022.

\bibitem{irs_loc_mag2}
K.~Keykhosravi, B.~Denis, G.~C. Alexandropoulos, Z.~S. He, A.~Albanese,
  V.~Sciancalepore, and H.~Wymeersch, ``{Leveraging RIS-enabled smart signal
  propagation for solving infeasible localization problems: Scenarios, key
  research directions, and open challenges},'' \emph{IEEE Veh. Technol. Mag.},
  {Early Access}.

\bibitem{emil22}
E.~Björnson, H.~Wymeersch, B.~Matthiesen, P.~Popovski, L.~Sanguinetti, and
  E.~de~Carvalho, ``{Reconfigurable intelligent surfaces: A signal processing
  perspective with wireless applications},'' \emph{IEEE Signal Process. Mag.},
  vol.~39, no.~2, pp. 135--158, Mar. 2022.

\bibitem{Dardari22}
D.~Dardari, N.~Decarli, A.~Guerra, and F.~Guidi, ``{LOS/NLOS near-field
  localization with a large reconfigurable intelligent surface},'' \emph{IEEE
  Trans. Wireless Commun.}, vol.~21, no.~6, pp. 4282--4294, Jun. 2022.

\bibitem{aoa}
W.~Wang and W.~Zhang, ``{Joint beam training and positioning for intelligent
  reflecting surfaces assisted millimeter wave communications},'' \emph{IEEE
  Trans. Wireless Commun.}, vol.~20, no.~10, pp. 6282--6297, Apr. 2021.

\bibitem{irs_aoa_toa1}
K.~Keykhosravi, M.~F. Keskin, G.~Seco-Granados, P.~Popovski, and H.~Wymeersch,
  ``{RIS-enabled SISO localization under user mobility and spatial-wideband
  effects},'' \emph{IEEE J. Sel. Topics Signal Process.}, vol.~16, no.~5, pp.
  1125--1140, Aug. 2022.

\bibitem{irs_aoa_toa2}
Y.~Han, S.~Jin, C.-K. Wen, and T.~Q.~S. Quek, ``{Localization and channel
  reconstruction for extra large RIS-assisted massive MIMO systems},''
  \emph{IEEE J. Sel. Topics in Signal Process.}, vol.~16, no.~5, pp.
  1011--1025, Aug. 2022.

\bibitem{irs_dvc1}
S.~Buzzi, E.~Grossi, M.~Lops, and L.~Venturino, ``{Foundations of MIMO radar
  detection aided by reconfigurable intelligent surfaces},'' \emph{IEEE Trans.
  Signal Process.}, vol.~70, pp. 1749--1763, Mar. 2022.

\bibitem{irs_dvc2}
H.~Zhang, H.~Zhang, B.~Di, K.~Bian, Z.~Han, and L.~Song, ``{MetaRadar:
  Multi-target detection for reconfigurable intelligent surface aided radar
  systems},'' \emph{IEEE Trans. Wireless Commun.}, vol.~21, no.~9, pp.
  6994--7010, Sep. 2022.

\bibitem{multiirs1}
Y.~Cheng, K.~H. Li, Y.~Liu, K.~C. Teh, and G.~K. Karagiannidis,
  ``{Non-orthogonal multiple access (NOMA) with multiple intelligent reflecting
  surfaces},'' \emph{IEEE Trans. Wireless Commun.}, vol.~20, no.~11, pp.
  7184--7195, May 2021.

\bibitem{multiirs3}
S.~Zhang and R.~Zhang, ``{Intelligent reflecting surface aided multi-user
  communication: Capacity region and deployment strategy},'' \emph{IEEE Trans.
  Commun.}, vol.~69, no.~9, pp. 5790--5806, Sep. 2021.

\bibitem{fdd}
N.~Saquib, E.~Hossain, and D.~I. Kim, ``{Fractional frequency reuse for
  interference management in LTE-advanced hetnets},'' \emph{IEEE Wireless
  Commun.}, vol.~20, no.~2, pp. 113--122, Apr. 2013.

\bibitem{Torrieri84}
D.~J. Torrieri, ``{Statistical theory of passive location systems},''
  \emph{IEEE Trans. Aerosp. and Electron. Syst.}, vol. AES-20, no.~2, pp.
  183--198, Mar. 1984.

\bibitem{Mao07}
G.~Mao, B.~Fidan, and B.~D.~O. Anderson, ``{Wireless sensor network
  localization techniques},'' \emph{Comput. Netw.}, vol.~51, no.~10, pp.
  2529--2553, Jul. 2007.

\bibitem{weighted_lasso}
N.~Vaswani and W.~Lu, ``{Modified-CS: Modifying compressive sensing for
  problems with partially known support},'' \emph{IEEE Trans. Signal Process.},
  vol.~58, no.~9, pp. 4595--4607, May 2010.

\bibitem{weighted_lasso_3}
M.~P. Friedlander, H.~Mansour, R.~Saab, and z.~Yilmaz, ``Recovering
  compressively sampled signals using partial support information,'' \emph{IEEE
  Trans. Inf. Theory}, vol.~58, no.~2, pp. 1122--1134, Feb. 2012.

\bibitem{weighted_lasso_4}
A.~Flinth, ``Optimal choice of weights for sparse recovery with prior
  information,'' \emph{IEEE Trans. Inf. Theory}, vol.~62, no.~7, pp.
  4276--4284, Jul. 2016.

\bibitem{sekino}
J.~Sekino, ``{n-Ellipses and the minimum distance sum problem},'' \emph{Amer.
  Math. Monthly}, vol. 106, no.~3, pp. 193--202, Mar. 1999.

\end{thebibliography}

\end{document}